\documentclass[useAMS,usenatbib,usegraphicx]{mn2e}

\usepackage[totalwidth=480pt, totalheight=680pt]{geometry}
\usepackage{epsfig,graphicx}
\usepackage{color}

\newcommand{\ergs}{ergs~s\ensuremath{^{-1}}}
\newcommand{\hal}{H\ensuremath{\alpha}}
\newcommand{\hbe}{H\ensuremath{\beta}}
\newcommand{\lhal}{\ensuremath{L_{\mathrm{H{\alpha}}}}}
\newcommand{\bhal}{\ensuremath{B_{\mathrm{H{\alpha}}}}}

\newcommand{\kms}{km~s\ensuremath{^{-1}}}
\newcommand{\per}{\ensuremath{^{-1}}}
\newcommand{\persq}{\ensuremath{^{-2}}}
\newcommand{\msun}{\ensuremath{M_{\sun}}}
\newcommand{\zsun}{\ensuremath{Z_{\odot}}}
\newcommand{\zoh}{\ensuremath{12+\log \mathrm{(O/H)}}}
\newcommand{\ahal}{\ensuremath{A_{\mathrm{H{\alpha}}}}}
\newcommand{\ahbe}{\ensuremath{A_{\mathrm{H{\beta}}}}}
\newcommand{\ebv}{\ensuremath{E\mathrm(B-V)}}
\newcommand{\debv}{\ensuremath{\Delta E\mathrm(B-V)}}
\newcommand{\hii}{H\,{\sc ii}}
\newcommand{\oiii}{[O\,{\sc iii}]}
\newcommand{\oii}{[O\,{\sc ii}]}
\newcommand{\oi}{[O\,{\sc i}]}

\newcommand{\nii}{[N\,{\sc ii}]}
\newcommand{\sii}{[S\,{\sc ii}]}

\title[Dust reddening in SF galaxies] {Dust reddening in star-forming galaxies}

\author[T. Xiao et al.]{
\parbox[t]{\textwidth}{
Ting Xiao\thanks{E-mail:xiaoting@mail.ustc.edu.cn (TX); twang@ustc.edu.cn (TW)},
Tinggui Wang\footnotemark[1],
Huiyuan Wang,
Hongyan Zhou,
Honglin Lu  \\
and Xiaobo Dong
}
\vspace*{6pt}\\
Key Laboratory for Research in Galaxies and Cosmology, \\
Department of Astronomy, University of Science \&
Technology of China, Chinese Academy of Sciences, Hefei, Anhui 230026, China}

\begin{document}

\date{Accepted . Received ; in original form }

\pagerange{\pageref{firstpage}--\pageref{lastpage}} \pubyear{}

\maketitle

\label{firstpage}

\begin{abstract}
We present empirical relations between the global dust reddening
and other physical galaxy properties including the \hal\ luminosity,
\hal\ surface brightness, metallicity and axial ratio for
star-forming disc galaxies.  The study is based on a large
sample of $\sim$ 22 000 well-defined star-forming galaxies selected
from the Sloan Digital Sky Survey (SDSS).  The reddening
parameterized by color excess \ebv\ is derived from the Balmer
decrement.  Besides the dependency of reddening on \hal\
luminosity / surface brightness and gas phase metallicity,
it is also correlated with the galaxy inclination, in
the sense that edge-on galaxies are more attenuated
than face-on galaxies at a give intrinsic luminosity.
In light of these correlations, we present the empirical formulae
of \ebv\ as a function of these galaxy properties, with a scatter
of only 0.07 mag.  The empirical relation can be reproduced if
most dust attenuation to the \hii\ region is due to diffuse
interstellar dust distributing in a disc thicker than that of \hii\ regions.
The empirical formulae can be incorporated into semi-analytical
models of galaxy formation and evolution to estimate the dust reddening and
enable comparison with observations more practically.
\end{abstract}


\section{Introduction}
Dust is a crucial component of galaxies in modifying the observed properties
of galaxies by absorbing and scattering starlight, and also re-emits the absorbed energy
in the mid- and far-infrared bands.  The extinction\footnote{The
nomenclature conventionally adopted is that, ``extinction'' describes the
absorption and scattering of light from a point source behind the dust screen;
the ``attenuation'' describes the reduction and reddening of the light in a
galaxy where stars and dust are mixed in geometrical distribution.  In this
work, we do not discriminate between these two terms when referring to the dust
attenuation within the galaxy, except for the Galactic ``extinction curve'' and
the ``attenuation curve'' for the starburst galaxies.  The extinction and reddening
can be converted to each other with an assumed extinction curve.} cross section
of dust generally decreases with increasing wavelength, i.e., the extinction is
more severe at shorter wavelength, particularly at ultra-violet (UV) band.
As a result, the spectral energy distributions (SED) of galaxies appear redder,
which is so-called ``reddening''.  The effects of dust should be properly accounted
for when interpreting the observations of the galaxies, for example the
luminosity or star formation rate (SFR) of a galaxy.  For a galaxy with spectroscopic
data covering suitable wavelengths, it is usually possible to estimate the amount of dust
extinction from the spectrum.  For example, the average dust reddening of \hii\ regions
can be derived from Balmer decrement, and be used to correct for the dust attenuation
to the emission-line luminosity.  In other cases, one has to rely on empirical relations
to get an estimate of dust extinction in a statistical way.

Additionally, it is necessary to incorporate dust into the galaxy formation
and evolution model.  Currently there are two ways to account for dust effects: full
radiative transfer calculations assuming dust properties and geometric distribution,
or adopting simple recipes obtained empirically.  Since the dust formation in the
galactic environment is a complicated chemical process, it has not yet been
implemented in the current galaxy evolution models.  The dust properties and
distribution are input in the models and should be tested with
observations.  Alternatively, many authors have adopted the empirical relations
to compute the perpendicular optical depth of a galactic disc, and then assign a random
inclination angle for each galaxy to get the final dust corrections (e.g.
Guiderdoni \& Rocca-Volmerange 1987; Kauffmann et al. 1999; Somerville \& Primack 1999;
De Lucia, Kauffmann \& White 2004; De Lucia \& Blaizot 2007; Kang et al. 2005;
Kitzbichler \& White 2007).  Therefore, a well-defined empirical recipe of dust
reddening will help the comparison of model predictions with observations.

Previous studies have suggested that dust reddening in star-forming galaxies is
correlated with the SFR, which can be well estimated with
the intrinsic \hal\ luminosity (e.g. Kennicutt 1998; Calzetti et al. 2010).
Galaxies with large SFR (or high luminosity) show strong global extinction in
the emission lines (e.g., Wang \& Heckman 1996; Charlot \& Fall 2000;  Calzetti
2001; Stasi\'nska \& Sodr\'e 2001; Afonso et al. 2003; Kewley et al. 2004;
Zoran, Barkana, \& Thompson 2006; Calzetti et al. 2007; Garn et al. 2010), and
also large infrared to ultraviolet flux ratios (e.g., Iglesias-Paramo et al. 2006
and reference therein).

In addition, it has been recognized that dust reddening might also be a function
of gas-phase metallicity, i.e. the reddening and extinction increase with
metallicity.  For instance, the average reddening to \hii\ regions in an
individual galaxy is correlated with the metallicity in the disc (e.g.,
Quillen \& Yukita 2001; Boisser et al. 2004).  For the case of whole galaxies,
the relationship between reddening and metallicity also exists (e.g. Heckman
et al. 1998; Buat et al. 2002; Asari et al. 2007).  It is also supported by
the fact that the low-metallicity galaxies such as blue compact galaxies are
usually less reddened (e.g. Kong 2004).

Recently, Garn \& Best (2010) also found a significant correlation between
dust extinction and metallicity, however, they claimed that the dependency
of reddening on stellar mass is more fundamental. They built a sample
of about 90 000 star-forming galaxies using Sloan Digital Sky Survey (SDSS)
data, and compare the relationship between the dust extinction of \hal\
luminosity and SFR, metallicity as well as stellar mass, respectively.  They
concluded that the dust extinction $\ahal$ can be best predicted from the
stellar mass, with a scatter of $\sim$0.3 mag.

Besides, the dependency of dust attenuation on inclination has been also
investigated.  While the correlation between dust attenuation of optical stellar
continuum and disc inclination has been well established for disc galaxies (e.g.
Driver et al. 2007; Shao et al. 2007; Unterborn \& Ryden 2008), that of emission
lines is still uncertain (e.g. Yip et al. 2010).

These works focused on how the dust reddening depends on one property of galaxies,
luminosity, metallicity or stellar mass, which are generally correlated
with each other (e.g. Garnett \& Shields 1987; Skillman et al. 1989; Zaritsky
et al. 1994; Zahid et al. 2011; Tremonti er al. 2004).  Disentangling the
dependencies of dust reddening on these galaxy properties can not only help us
to derive a more accurate empirical formulae for dust reddening, but also provide
more insight to the dust formation within galaxy environment.  In this work, we study
a large sample of star-forming disc galaxies in the local universe with a median
redshift of 0.07, selected from the spectroscopy database of SDSS.
We obtain empirical formulae for dust reddening as a function of
intrinsic \hal\ luminosity / surface brightness, gas-phase metallicity and the axial
ratio of the disc.

In Section~\ref{sec:sample} we describe our sample selection, and in
Section~\ref{sec:methods} describe the spectral analysis and methods used to
estimate the parameters such as the dust reddening, intrinsic \hal\ luminosity,
surface brightness, metallicity and axial ratio etc.  We present the resultant
expression for the empirical formulae of dust reddening, and compare our results
with previous work in Section~\ref{sec:relation}.  In Section~\ref{sec:toymodel} we
reproduce the observed trend with a toy model of parallel-slab disc, which gives
some insights into the geometry of dust distribution and dust-to-gas ratio.
Conclusions are given in Section~\ref{sec:conclusion}.  Throughout the paper,
we will assume a cosmology of $H_0=72$ \kms Mpc\per, $\Omega_M=0.3$ and $\Omega_\Lambda=0.7$.

\section{DATA AND SAMPLE SELECTION}
\label{sec:sample}

We start from the spectroscopic sample of galaxies in the SDSS data release five
(DR5, Adelman-McCarthy et al. 2007).  SDSS produces imaging and spectroscopic
survey with a wide-field 2.5m telescope at Apache Point Observatory, New Mexico (York et al. 2000).
The survey provides imaging in five broad bands \textsl{u, g, r, i, z}, with magnitude
limits of 22.2 in \textsl{r} band, and spectroscopic targets are selected using a variety
of algorithms, including the ``main'' sample of galaxies with $r$-band magnitude
brighter than 17.77 (Strauss et al. 2002) with fibers of 3\farcs\ diameter.
The spectra range from 3200 to 9200 \AA\ at a resolution $R\equiv\lambda/\Delta \lambda=1800$ \footnote{www.sdss.org}.  DR5 spectroscopic area covers 5740 deg\persq, and there are
$\sim$675,000 spectra classified as galaxies.  To select clean galaxies, we use the SDSS
photometric flags to eliminate the targets that are only one part of a large galaxy or a
part of merging galaxies.  We list the sample-selection cuts and corresponding
number of remaining objects within each subsample in Table~\ref{tab:sample}.
Then the galaxy spectra are corrected for the
Galactic extinction using the dust extinction map (Schlegel et al. 1998)
with an extinction curve of Fitzpatrick (1999) with $R_V=3.1$.

We select emission-line galaxies where the \hal\ has been
detected at high significance, i.e., S/N$>$20.  To include galaxies of high
reddening, we use a looser criterion on \hbe: S/N$>$10.  The emission
lines are measured with the procedures described in details in Section~\ref{sec:specfit}.
In order to use the ratios of \oiii/\hbe\ and \nii/\hal\ on line-ratio
diagnostic diagrams (e.g. BPT diagrams; Baldwin, Phillips \& Terlevich 1981)
for spectral classification, we also impose
the criteria \oiii\ S/N$>$10 and \nii\ S/N$>$5.  It should be noted that
the requirement of lower limit in the S/N ratio for \nii\ may drop some low
metallicity galaxies, and that for \oiii\ may miss high metallicity galaxies.
We will justify the S/N criteria in Section~\ref{sec:erranaly}.

Galaxies with significant broad \hal\ component are rejected.  The broad-line
Active Galactic Nuclei (AGNs) are defined as objects for which adding an
additional broad component of \hal\ to the emission-line model can
significantly improve the fit to the \hal+\nii\ blend (refer to Dong
et al. 2005, 2007; also Zhou et al. 2006).  In practice, the galaxies with
broad \hal\ component detected at the $\ge 5\sigma$ significance level
are regarded as candidates of broad-line AGNs, and removed from the sample.
We also remove the narrow-line active galaxies by using the BPT diagram (Kauffmann
et al. 2003b, hereafter Ka03; Kewley et al. 2006) based on \nii/\hal.
The galaxies below the Ka03 pure star-formation line on \nii/\hal\ diagram
are referred as our sample of star-forming galaxies.  Most of these galaxies
lie below the extreme-starburst line on the \sii/\hal\ and \oi/\hal\
diagrams, with a fraction of 99 percent and 94 percent, respectively.  Using a
more strict criterion on selecting \hii\ galaxies given by Stasi{\'n}ska et al.
(2006) will result in less metal-rich galaxies (see \S~\ref{sec:metal}).

As shown by Kewley et al. (2005), if the nuclear spectrum contains less than
twenty percent of the total galaxy light, we will likely over-estimate
the global metallicity and reddening, and under-estimate the global SFR
by a significant fraction.  Therefore, we remove the galaxies for which
the fiber magnitude at $g$-band is greater than the total magnitude by 1.7 mag.

In the following analysis, we will pick out the late-type (presumably
disc-dominated) galaxies from the star-forming galaxies based on the
likelihoods provided by SDSS pipeline.  The photometry pipeline provides
the likelihoods (dev\_L, exp\_L, and star\_L)
associates with the de Vaucouleurs, exponential, and PSF fits, respectively.
The fractional likelihoods for the exponential fit is calculated as
\begin{equation}
f(\mathrm{exp}\_\mathrm{L}) = \frac{\mathrm{exp}\_\mathrm{L}}{\mathrm{exp}\_\mathrm{L} + \mathrm{dev}\_\mathrm{L} + \mathrm{star}\_\mathrm{L}} ,
\label{eq:frac-lkl}
\end{equation}
and similarly for $f$(exp\_L) and $f$(star\_L).  It is suggested that
the fractional likelihood greater than 0.5 for any of the three model
fits is generally good as a threshold for object classification (Stoughton et al. 2002).
For a galaxy, $f$(star\_L) is generally zero.  We define a galaxy as a disc
galaxy if the logarithmic likelihood for an exponential fit ($r$-band)
is larger than that of a de Vaucouleurs fit by 0.2 dex (corresponding to
$f$(exp\_L)$>0.6$), while as an elliptical galaxy if the logarithmic
likelihood for a de Vaucouleurs fit is larger than that of an exponential
fit by 0.2 dex ($f$(dev\_L)$>0.6$).  This parameter is also correlated with the compactness
index such as $R_{50}/R_{90}$, which is commonly used in quantitative
classification (e.g. Shimasaku et al. 2001; Strateva et al. 2001).
With this criterion, 23919 star-forming galaxies are classified as disc
galaxies, and 7650 as elliptical galaxies, and 595 as unclassified type.

\setcounter{table}{0}
\begin{table*}
 \centering
 \begin{minipage}{120mm}
  \caption{Summary of sample-selection cuts.  }
  \label{tab:sample}
  \begin{tabular}{@{}lrr@{}}
  \hline
  Selection cut & Num remained &  Percent removed by cut  \\
 \hline
SDSS DR5 spectroscopic sample                   & 582 512 &    /    \\
Photometric clean sample                        & 495 165 &   14.99 \\
S/N(\hal) $>$ 20                                & 191 289 &   61.37 \\
S/N(\hbe) $>$ 10                                & 128 046 &   33.06 \\
S/N(\oiii) $>$ 10                               &  67 236 &   47.49 \\
S/N(\nii) $>$ 5                                 &  67 013 &    0.33 \\
Remove broad-line AGN candidates                &  57 068 &   14.84 \\
Remove multiple observations                    &  56 241 &    1.45 \\
Select star-forming galaxies using BPT          &  46 865 &   16.67 \\
Require $>$20\% total light in the fiber        &  32 164 &   31.37 \\
Select disk galaxies                            &  23 919 &   25.63 \\
Require $-2.5<N2<-0.4$                          &  22 616 &    5.45 \\
 \hline
\end{tabular}
\end{minipage}
\end{table*}

\section{METHODS AND SAMPLE PROPERTIES}
\label{sec:methods}

\subsection{SPECTRA ANALYSIS: STARLIGHT-CONTINUUM SUBTRACTION AND EMISSION-LINE FITTING}
\label{sec:specfit}

In order to measure the emission lines, we take two steps to analyze
the spectra: continuum fitting and emission-line fitting.  First, we
subtract the stellar continuum following the recipe described by Lu et al. (2006).  In brief,
Ensemble Learning for Independent Component Analysis (EL-ICA) has been applied
to the simple stellar population library (Bruzual \& Charlot 2003, BC03) to
derive a set of templates, which then are shifted and broadened to match
the stellar velocity dispersion of the galaxy, and reddened assuming a
starburst-like extinction law to fit the observed galaxy spectra.  During the
continuum fitting the bad pixels flagged out by SDSS pipeline as well as the
emission-line regions are masked.  From the fit, we obtain simultaneously the
modeled stellar-light component, stellar velocity dispersion and an effective
reddening\footnote{The effective reddening derived in this process is fairly
well correlated with that of emission lines estimated using the Balmer
decrements for \hii\ galaxies (see also e.g. Calzetti et al. 1994; Stasi{\'n}ska et al.
2004).} to the stellar light (refer to Lu et al. 2006 for details).
Second, the emission lines are modeled with various Gaussians on the
continuum-subtracted spectra, using the \texttt{MPFIT} package (Markwardt 2009)\footnote{
\texttt{MPFIT} package includes routines to perform non-linear least squares curve
fitting, kindly provided by Craig B. Markwardt, available at
http://purl.com/net/mpfit.} implemented in Interactive Data Language (IDL).
The formal 1$\sigma$ errors in flux obtained from the fitting, are propagated
from the error of the spectra, and then adopted as the emission-lines
flux uncertainties.  The emission lines we measure have been corrected for
absorption lines, by subtracting the stellar component models.  The typical
absorption correction is 27 percent of the flux of \hbe\ emission-line.
We examine the model-fitting of higher-order Balmer absorption lines, like H$\delta$,
which is less contaminated by emission lines, and estimate the uncertainty
in absorption measurement to be less than 24 percent, including the statistical
uncertainty.  Thus the uncertainty in \hbe\ absorption measurement is generally
less than 6 percent of \hbe\ emission flux.

The corresponding emission-line regions
are masked in the continuum fits.  To determine the proper mask-ranges
for the emission lines, usually several iterations of the above procedures
are required.  Emission lines, \hal, \hbe, H$\gamma$, H$\delta$, H$\epsilon$,
\oii $\lambda 3727$ \AA, [Ne\,{\sc iii}] $\lambda\lambda 3869,3969$ \AA\AA,
[Ne\,{\sc v}] $\lambda 3427$ \AA, \oiii $\lambda 4363$ \AA, [He\,{\sc ii}]
$\lambda 4686$ \AA, \oiii $\lambda\lambda 4959,5007$ \AA\AA, [N\,{\sc i}]
$\lambda 5199$ \AA, [He\,{\sc i}] $\lambda 5876$ \AA, \nii $\lambda\lambda
6548,6584$ \AA\AA, \sii $\lambda\lambda 6716, 6731$ \AA\AA, \oi $\lambda\lambda
6300,6344$ \AA\AA, [Ar\,{\sc iii}] $\lambda 7136$ \AA\ are included in the
first iteration, but insignificant ones (signal-to-noise ratio in
emission-line flux, S/N$<$3) are dropped in the later fitting.  For robustness,
we assume identical profiles for \nii\ doublet lines and \hal.  The \sii\ doublet
lines are assumed to have a same profile, so are \oiii\ doublet lines.  The ratios
of \nii\ doublets and \oiii\ doublets are fixed to their theoretical values, 2.96 and 3,
respectively.  \oii\ doublets are each modeled with a single gaussian of the same width.
To reduce the uncertainty in the measurements of weak emission lines, we fix
their profiles to those of strong lines of similar ionization states.  As the
final procedure, upper limits are given to the undetected lines assuming that
the lines have the identical profile as the detected strong lines.  If no
emission-line has been detected significantly ($>5\sigma$), no emission-line flux
will be given for that spectrum.

\subsection{REDDENING AND \lhal\ CORRECTION}
\label{sec:lhacor}

\begin{figure*}
\includegraphics[width=0.8\textwidth]{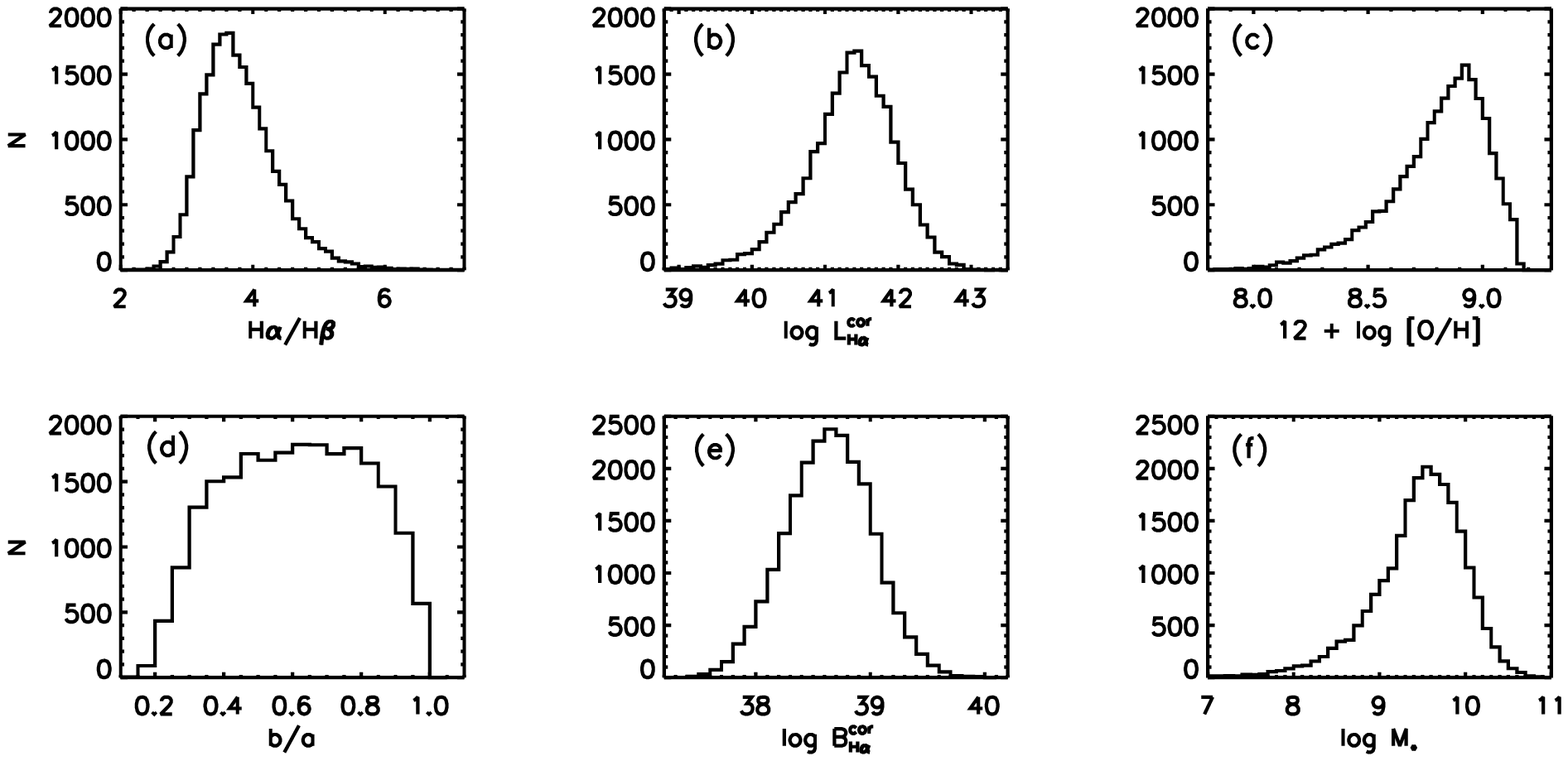} \caption{Distributions of properties for the
selected $\sim$22 000 disc-dominated \hii\ galaxies in the SDSS DR5 (see text): (a) Balmer
Decrement \hal/\hbe; (b) dust-attenuation corrected \hal\ luminosity
$\log\ L_{\mathrm{H\alpha}}^{cor}$; (c) \zoh; (d) axial ratio {\em b/a}; (e) \hal\ surface
brightness corrected for dust attenuation $\log\ B_{\mathrm{H\alpha}}^{cor}$ , see
Equation~\ref{eq:br}; (f) stellar mass.
 } \label{fig:properties}
\end{figure*}

We estimate the reddening of emission lines using the Balmer decrement
\hal/\hbe\ ratio.  An intrinsic value of 2.86 as expected for case-B
recombination with electron density $n_e=100{\rm cm}^{-3}$ at
$T=10^4$ K is assumed (Osterbrock \& Ferland 2006).
This value is generally consistent with the lower limit of the measured
\hal/\hbe\ ratio (Figure~\ref{fig:properties}a) in our sample.
There is only a small fraction (about 1.5 percent) of objects
with \hal/\hbe\ below 2.86, likely due to measurement uncertainty.
For these objects, the reddening is adopted as zero.
Due to our stringent criteria for \hbe\ and \oiii\ detections, a significant
fraction of objects with high Balmer decrement values have been dropped;
thus most of our sample have \hal/\hbe\ $<7$.  The attenuation $A_{\rm \hal}$
is estimated from the Balmer decrement \hal/\hbe\ assuming an
extinction curve, and then used to correct \hal\ luminosity.

We also apply aperture correction on \hal\ luminosity, based on
the difference between the model magnitude and the fiber magnitude
at $g$-band.  This correction method assumes the distribution of
\hal\ emission is the same as that of the stellar light (continuum emission).
Such a correction is only an approximation because the line emission and
the stellar continuum emission may not distribute in the same way.
Some other authors use empirical approach to make aperture corrections
taking into account the color differences within / outside the
fiber (Brinchmann et al. 2004), or constrain the global SFR from
fitting stochastic models to the photometric SED (Salim et al. 2007).
The former method is based on the main assumption that the distribution of specific SFR
for a given set of colours inside the fibre is similar to that outside.  We test
with this method, but find that at a given set of colours the likelihood distribution
of specific SFR inside the fiber varies with Balmer decrement.  The typical specific SFR
is higher for galaxies with larger Balmer decrement, and lower for galaxies with smaller Balmer
decrement.  Thus this method of aperture correction may introduce dependence of \hal\
luminosity on Balmer decrement.  Since our goal is to investigate the correlation
between dust reddening and luminosity, we decide to settle for the simple scaling
method.  We also examine if the global SFR obtained with
the method of Salim et al. (2007) is used, and compare with the SFR estimated
from far-infrared luminosity, as we will check for our corrected SFR in the following
part of this section.  The test suggests our simple method is no worse than theirs.

Because different parts of a galaxy suffer from different extinction, the extinction
derived from the Balmer decrement is only a certain average.  In the following, we
will check if the aperture and attenuation correction introduce any fake correlation between
the \hal\ luminosity and reddening.  If the corrected $f$(\hal) does not represent an
accurate intrinsic \hal\ flux, then the star-formation rate
estimated from \hal\ luminosity will be inaccurate.  To examine this issue,
we use far-infrared luminosity as the reference tracer for SFR and compare
it with the SFR estimated from \hal\ luminosity.  On one hand, although
integrated IR emission should provide a robust measurement of SFR in
dusty circumstances (Kennicutt 1998 and references therein, Dale \&
Helou 2002), there are also calibrations based on luminosities of
specific bands at infrared (e.g. Wu et al. 2005; Alonso-Herrero et
al. 2006; Zhu et al. 2008; Calzetti et al. 2007, 2010; Rieke et al.
2009).  At high luminosity ($L_{\rm 70\mu
m}\ga1.4\times10^{42}$\ergs), $L_{\rm 70\mu m}$ correlates linearly
with SFR, thus could be used as a tracer of SFR (Calzetti et al.
2010, their equation 22):
\begin{equation}
SFR(70)\ (\msun {\rm yr\per}) = 5.88\times10^{-44}L_{\rm 70\mu m}.
\label{eq:sfr-l70}
\end{equation}
On the other hand, we adopt the calibration of Calzetti et al. (2010, their equation
5) to convert \hal\ luminosity to SFR as:
\begin{equation}
SFR({\rm \hal})\ (\msun {\rm yr\per}) = 5.45\times10^{-42}\lhal(\rm
\ergs), \label{eq:sfr-lha}
\end{equation}
in which \lhal\ should be corrected for intrinsic extinction.
This calibration is based on solar metallicity and Kroupa (2001)
Initial Mass Function (IMF).  The Kroupa IMF has two power laws,
one with a slope of $-1.3$ for stellar masses ranging from
0.1 to 0.5 \msun\ and the other with a slope of $-2.3$ for stellar masses
ranging from 0.5 to 100 \msun. This calibration is based on a $t>1$ Gyr age
constant star-formation stellar population.

\begin{figure*}
\setlength{\unitlength}{.5in}
\begin{picture}(14,4)
\put(0, 0){\includegraphics{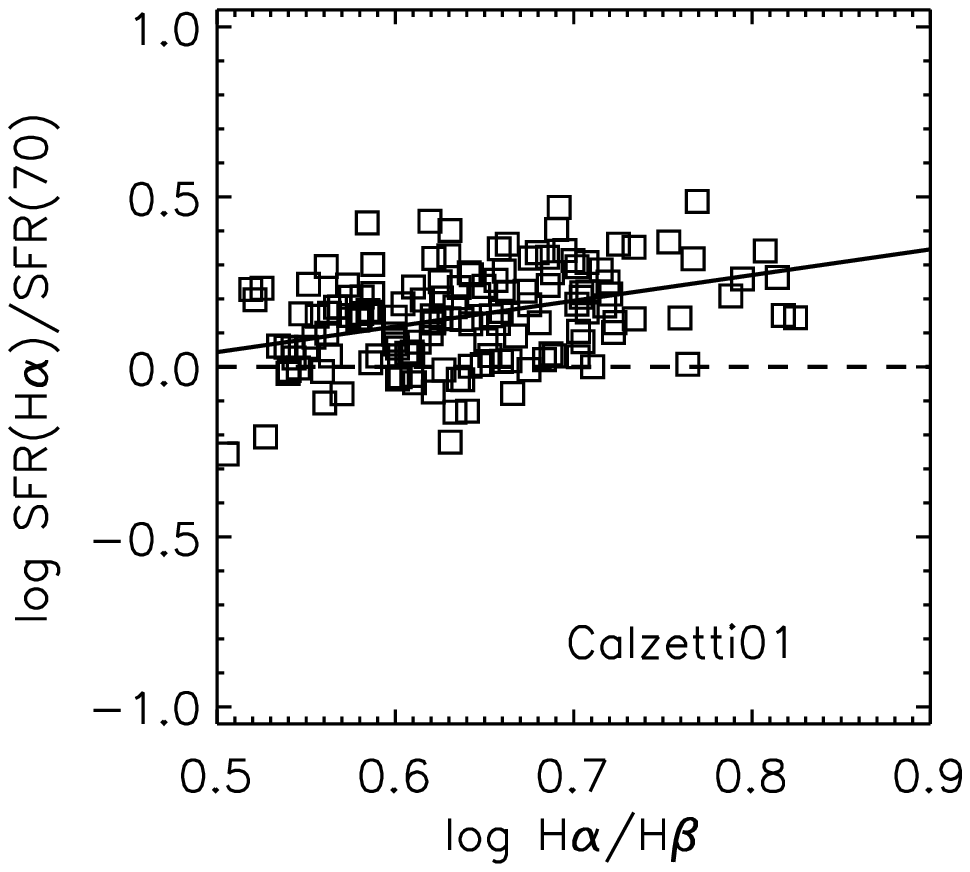}} \put(4, 0){\includegraphics{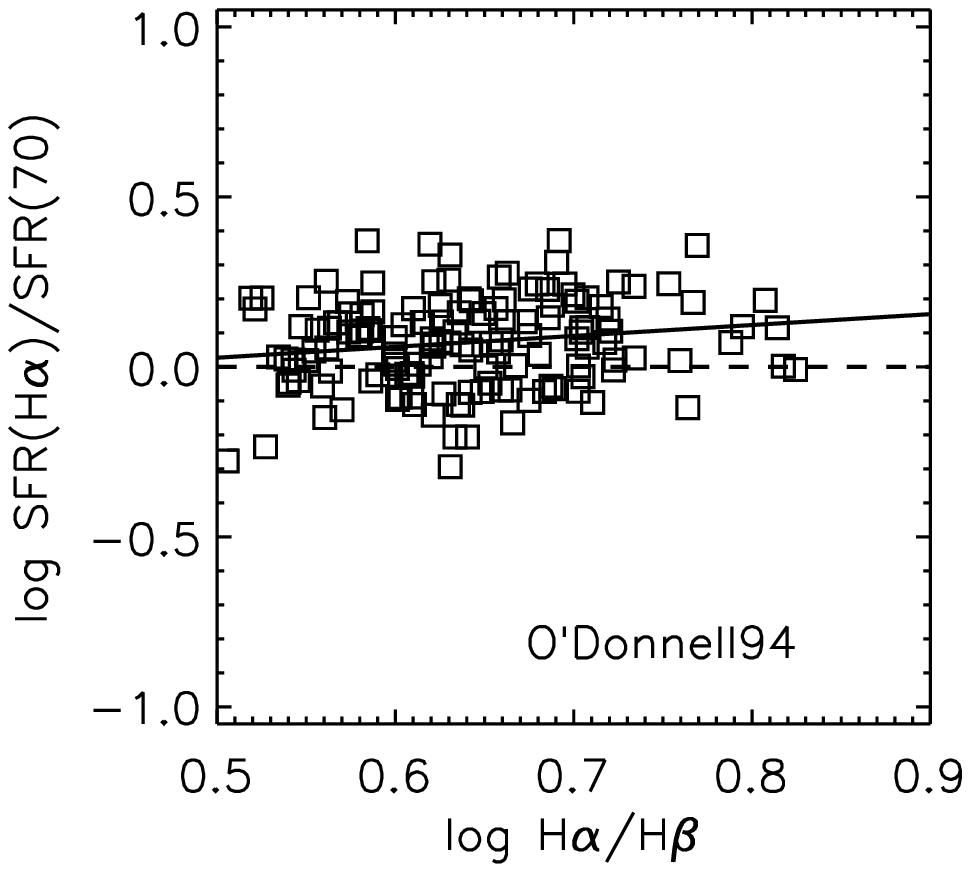}} \put(8,
0){\includegraphics{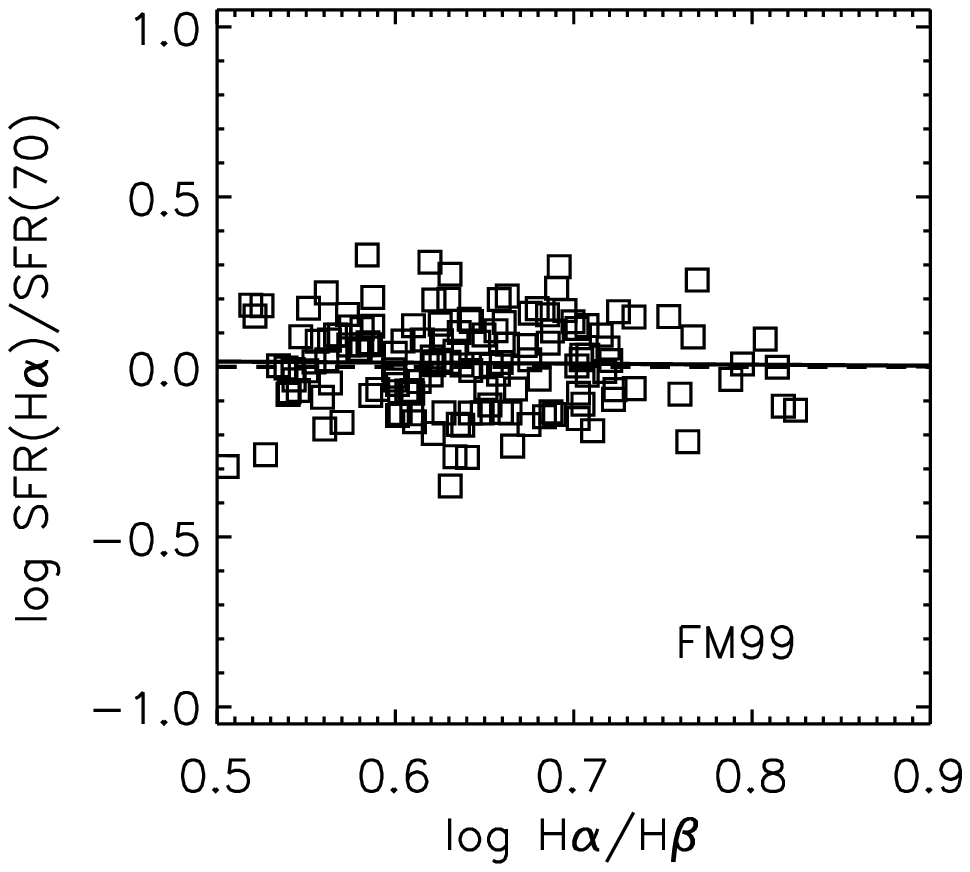}}
\end{picture}
\caption{The ratio of ${\rm SFR}({\rm \hal})/{\rm SFR}(70)$ versus Balmer decrement
\hal/\hbe\ in logarithmic scale.  ${\rm SFR}({\rm \hal})\ (\msun {\rm yr\per}) =
5.45\times10^{-42}\lhal(\rm \ergs)$, where \lhal\ is
attenuation-corrected with a starburst attenuation curve (Calzetti
2001, left panel), or a Galactic extinction curve (middle panel:
O'Donnell 1994 curve, right panel: Fitzpatrick 1999 curve).
${\rm SFR}(70)\ (\msun {\rm yr\per}) = 5.88\times10^{-44}L_{\rm 70\mu m}$.
Solid lines represent the best-fitting for the sample, and dashed lines
represent ${\rm SFR}({\rm \hal})/{\rm SFR}(70)=1$.  } \label{fig:extcur}
\end{figure*}

We cross-match our parent sample of star-forming galaxies with the 70$\mu$m
band photometry catalogs from Spitzer Wide-area InfraRed
Extragalactic Survey (SWIRE; Lonsdale et al. 2003) Data Release 3.  To
estimate the \hal\ SFR for the galaxies with confidence, we require the
aperture to include at least twenty percent of the total light (Kewley et al.
2005). We also require the detection of \hal\ emission to be more significant
than $10\sigma$.  In order to get a matched sample of reasonable size, we
apply a looser S/N criteria on other emission lines (i.e. S/N$>$10 for \hal,
and S/N$>$5 for \nii, \hbe, \oiii).  With a matching radius of 5\arcsec, we
get 156 star-forming galaxies with measurements of 70$\mu$m fluxes.  Removing
five galaxies with contaminating sources nearby in IR emission, two
galaxies with unreliable or problematic 70$\mu$m fluxes, and one galaxy with
problematic aperture correction (negative value), there are 148
galaxies left in the SWIRE star-forming galaxy sample.

We correct \lhal\ for dust attenuation with three extinction curves
respectively: 1) the attenuation curve for the continuum of starburst
galaxies (Calzetti et al. 2000); and Galactic extinction curve of 2)
O'Donnell (1994); or 3) Fitzpatrick (1999).  Converting \lhal\ to SFR
with Equation~\ref{eq:sfr-lha}, we investigate the ratio of
SFR(\hal)/SFR(70) as a function of the Balmer decrement in the
log-space.  If \lhal\ is properly corrected for dust attenuation,
the ratio of $\lhal/L_{\rm 70\mu m}$, thus the ratio of
SFR(\hal)/SFR(70) should be independent of the dust reddening
and the Balmer decrement.  In the SWIRE star-forming galaxy sample,
there are 147 galaxies with $L_{\rm 70\mu m}$ greater than
$1.4\times10^{42}$ \ergs\ $\sim3.7\times10^8$ L$_{\sun}$, for which
Equation~\ref{eq:sfr-l70} can be used to estimate ${\rm SFR}(70)$.
Figure~\ref{fig:extcur} shows that if the
starburst attenuation-law is adopted (left panel), SFR(\hal)/SFR(70)
still correlates with $\hal/\hbe$ (the spearman rank coefficient $r_s=0.33$,
with the probability of null hypothesis $P_{null}<10^{-4}$), implying that the
\lhal\ might have been over-corrected than demanded, while the Galactic
extinction-curves give better correction on average (middle and
right panel).  With Fitzpatrick (1999) curve, the ratio ${\rm SFR}({\rm
\hal})/SFR(70)$ is uncorrelated to the Balmer decrement ($r_s=-0.01$,
$P_{null}=0.90$).  This means the \hal\ SFR can be well determined
from the attenuation-corrected \lhal\ by assuming Fitzpatrick's curve.
Therefore, we will adopt this curve for attenuation correction to
\lhal\ in the following analysis.

For our final sample of star-forming disc galaxies, the corrected \lhal\ is
shown in distribution histogram in Figure~\ref{fig:properties}(b).
The dust-extinction corrected \hal\ luminosity is in the range from 4$\times 10^{37}$ \ergs
to 2$\times 10^{43}$ \ergs, with a median value of 3$\times 10^{41}$ \ergs.  The
typical error in \lhal\ is 0.04 dex, as the quadrature sum of the measurement uncertainty
in observed flux of \hal\ emission, and the Balmer decrement used for attenuation correction.
The uncertainty induced by either the average reddening we simply assumed or the aperture
correction for \lhal\ has not been included.
Assuming Fitzpatrick's extinction curve, the color excess is estimated as
\setlength\arraycolsep{0.5mm}
\begin{eqnarray}
\ebv\ = & 1.99 \log \frac{\hal/\hbe}{2.86}   & for \; {\rm \hal/\hbe} \geq 2.86 \nonumber \\
      = & 0                                  & for \; {\rm \hal/\hbe} < 2.86 .
\label{eq:ebv}
\end{eqnarray}
The median formal uncertainty in \hal/\hbe\ for our selected
sample is typically 3.4 percent, which gives an uncertainty of 0.03~mag in
\ebv.  The uncertainty of \ebv\ is about 0.04 mag typically for the small sample of
SWIRE star-forming galaxies.  We also calculate \ebv\ from the ratio of \hal/H$_\gamma$
assuming Fitzpatrick's extinction curve, and the derived \ebv\ is quite consistent
with the value obtained using Equation~\ref{eq:ebv}.  This proves that our corrections
for Balmer absorptions are quite robust.

We note that in the relation between SFR(\hal)/SFR(70) and the
Balmer decrement, irrespective of which extinction curve adopted,
there is a moderate scatter $\sim$0.13 dex.  This scatter includes
the measurement errors of observed \lhal, the Balmer decrement and model/fiber
magnitudes at $g$-band;  the uncertainty in aperture-correction
and attenuation-correction; the measurement error in observed $L_{\rm 70\mu m}$;
and the calibration error in the SFR(\hal)/SFR(70) ratio.
The overall measurement error for SFR(\hal) is typically 0.04 dex, and for SFR(70)
typically 0.01 dex.  Then the remaining scatter of 0.12 dex accounts for the sum
in quadrature of uncertainties in aperture and attenuation correction, and the
the calibration scatter in SFR(\hal)/SFR(70).  Therefore, any one of these
uncertainties, for example the the calibration scatter in SFR(\hal)/SFR(70) should
be less than 0.12 dex, which is smaller than the calibration uncertainty in SFR(70)
of Equation~\ref{eq:sfr-l70}, $\sim$0.2 dex (Calzetti et al. 2010, refer to their Section 5).
Note in passing, we obtained a larger scatter ($\sim$0.17dex) in SFR(\hal)/SFR(70) if
MPA-JHU\footnote{MPA-JHU catalogue and the total SFRs are provided on the website
http://www.mpa-garching.mpg.de/SDSS/DR7/.} SFR(\hal) is used instead.

\begin{figure*}
\includegraphics[width=0.8\textwidth]{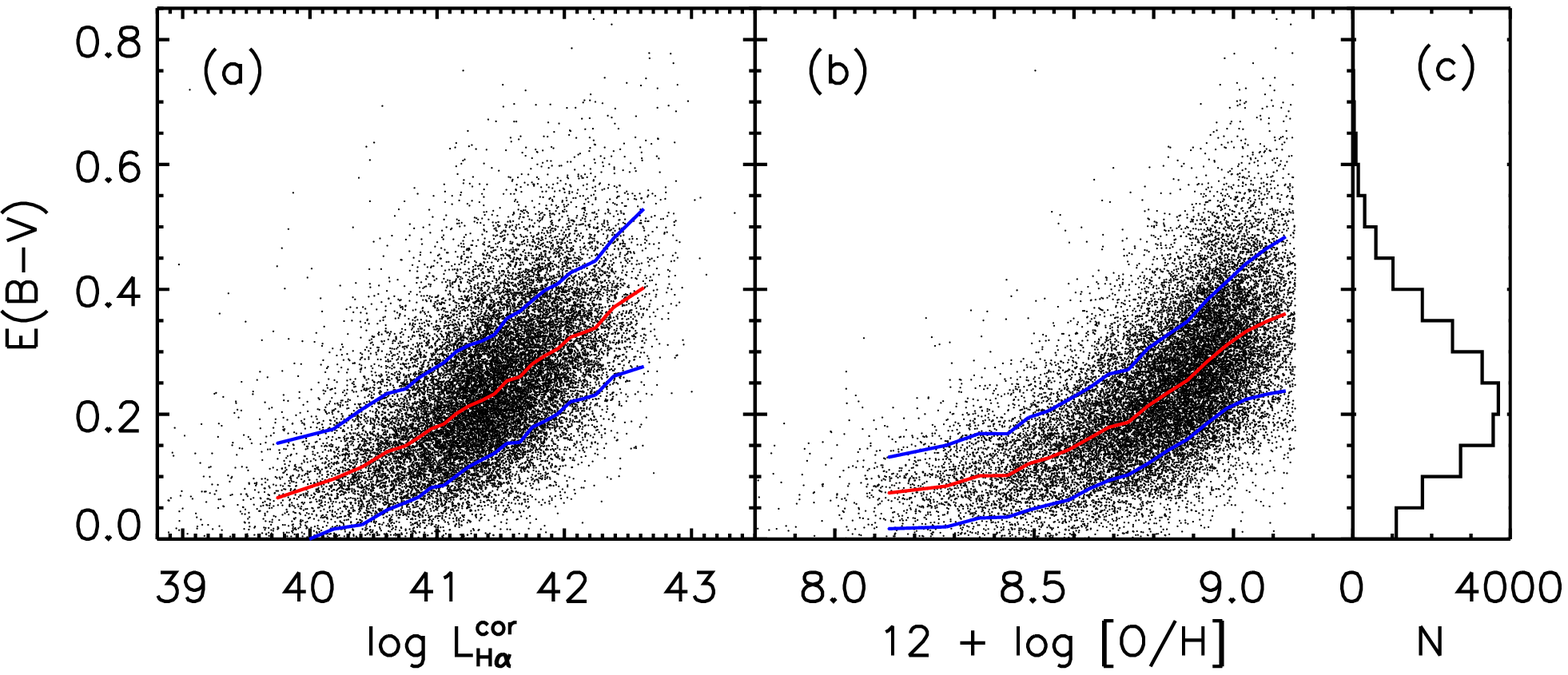}
\caption{The reddening of emission lines
parameterized by \ebv\ as a function of (a) attenuation-corrected
\hal\ luminosity \lhal\ (in units of \ergs), (b) oxygen abundance
for our final sample of 22616 star-forming disc-dominated galaxies.
The red line is the median value, while the blue lines enclose
68 percent (1$\sigma$) objects at a given $\log\ L_{\mathrm{H\alpha}}^{cor}$ or \zoh.
(c) Distribution of \ebv.
 }  \label{fig:ebv-lha-z}
\end{figure*}

\subsection{METALLICITY}
\label{sec:metal}

Estimating global metallicity has been widely studied by using strong lines.
However, there is still no consensus on which line ratio should be used.
Ratios of strong lines, such as $R_{23} \equiv \log$(\oii $\lambda 3727$ \AA\
+ \oiii $\lambda\lambda 4959, 5007$ \AA \AA) / \hbe,
\nii $\lambda 6584$ \AA\ / \oii $\lambda 3727$ \AA,
$N2 \equiv \log$(\nii $\lambda 6584$ \AA\ / \hal), and
$O3N2 \equiv \log$(\oiii $\lambda 5007$ \AA\ / \hbe) / (\nii $\lambda 6584$ \AA\ / \hal)
are used to estimate the oxygen abundance (e.g., McGaugh 1991;
Storchi-Bergmann et al. 1994; Denicol{\'o} et al. 2002;
Kewley \& Dopita 2002; Pilyugin 2003; Pilyugin \& Thuan 2005;
Pettini \& Pagel 2004; Tremonti et al. 2004; Liang et
al. 2006; Shi et al. 2006; Nagao et al. 2006).  The commonly used
metallicity indicator $R_{23}$ is not used here, because it includes
\oii\ emission line which is much prone to dust extinction.  Note that
one purpose of our work is to investigate the
relation between dust reddening and the metallicity, we should avoid any
spurious correlation potentially caused by the reddening correction. The estimator $N2$ and
$O3N2$ are both not sensitive to the reddening correction.  But the measurement
error of $O3N2$ (typically 0.03 dex) is larger than that of $N2$ (typically
0.01 dex).  Therefore $N2$ is preferred as our metallicity diagnostic.

The relation of line ratio versus abundance can be generally
calibrated using two different approaches: the empirical method relying on the
electron temperature $T_{e}$ as a surrogate for metallicity (cooling
increases with metallicity), or a comparison with photoionization
models.  Note that different methods lead to systematic differences in the
calibrated relations up to a factor of three or more in some extreme
cases, apart from the limited range for the validation of the
relations (refer to Kewley et al. 2008 for a detailed discussion).

We estimate the metal abundance with $N2$ index of Pettini \& Pagel (2004,
hereafter PP04), further revised by Nagao et al. (2006) and Liang et al. (2006).
PP04 derived their formula from \hii\ regions with well determined oxygen
abundance based on $T_{e}$-method, while Nagao et al. and Liang et al.
calibrated gas metallicity in galaxies obtained using other methods
such as $T_{e}$-method or Bayesian-technique by comparing with the
theoretical models of Tremonti et al. (2004). The three calibrations
yield very similar estimates of [O/H] at low abundances (e.g. $\zoh<8.5$),
but they deviates considerably at high abundances (e.g. $\zoh>8.5$): PP04
gives a lower abundance value than those of Liang et al. (2006) and
Nagao et al. (2006). Since PP04 calibration does not extend to high abundance (most of their
objects have $7.5<\zoh<8.5$), we adopt the calibration of Nagao et al. (2006),
expressed as
\begin{equation}
N2 = 96.641-39.941\times y+5.2227\times y^2-0.22040\times y^3,
\label{eq:n2}
\end{equation}
where $y\equiv\zoh$.  Equation~\ref{eq:n2} is valid within the metallicity
range $7.15\leq\zoh\leq9.16$, corresponding to $-2.5\leq N2\leq
-0.4$.\footnote{Nagao et al. (2006) did not specify the valid range of $N2$ for
their calibration, we adopt the range of their observational sample used for
calibration.  We are prudent to not use the extrapolation when $N2>-0.4$,
which saturates and yields quite high and unreasonable metallicities.}
We estimate metallicity for our sample with $N2$ for consistency, thus we
reject those with $N2> -0.4$ (only about five percent in our sample).  Note
that this selection has little effect on the results we obtain in this
paper.  The final sample consists of 22616 galaxies, with the oxygen abundance
in the range of $7.64<\zoh<9.16$ and with a median value $<\zoh>=8.84$ (Figure~\ref{fig:properties}c).
The oxygen abundances could be converted to metallicity in units of solar
metallicities (\zsun=0.02), adopting a value of $\zoh_{\odot}=8.66$
(Asplund et al. 2004), and the range is $0.1<Z/Z_{\odot}<3.2$, with a median
of $<Z/Z_{\odot}>=1.5$.

The error in the metallicity includes measurement
uncertainty and calibration uncertainty.  For our sample, the metallicity
is estimated using $N2$, and the typical measurement uncertainty in $N2$
is 2.5 percent, which corresponds only up to 0.02 dex in \zoh.  The main
uncertainty in metallicity estimation comes from the scatter of the
calibration of $N2$ indicator.  Nagao et al. did not explicitly provide the
scatter of their calibration (Equation~\ref{eq:n2}).  However, PP04 proposed
a linear calibration of $N2$ and indicated the $1\sigma$ uncertainty in $Z$
is 0.18 dex.  The $N2$ calibration in Equation~\ref{eq:n2}
differs significantly ($\Delta Z \geq 0.2$ dex) from PP04's calibration
only at metallicities $\zoh<7.5$ and $\zoh>8.5$.  Moreover, Kewley \& Ellison (2008)
compared various metallicity calibrations, and concluded that metallicities estimated
from strong-line methods should be consistent within 0.15 dex.  Therefore, we estimate
the upper limit of the uncertainty in metallicity to be $\Delta Z \sim 0.15$ dex.

\subsection{SURFACE BRIGHTNESS \bhal, DISC INCLINATION AND OTHER PROPERTIES}

The surface brightness of a galaxy is the flux received from a unit solid angle
as it appears on the sky, then we define the intrinsic surface brightness \bhal\
as the \hal\ luminosity per area within the half-light radius of the galaxy, i.e.
\begin{eqnarray}
\bhal \equiv \frac{\lhal}{8\pi^2 R_{50}^2} \quad\  ({\rm \ergs kpc\persq}),
\label{eq:br}
\end{eqnarray}
where $R_{50}$ is the Petrosian half-light physical radius (in kpc) in $r$-band, which
can be obtained by the product of Petrosian half-light radius (in arcsec) provided by
the SDSS pipeline, and the angular distance.  The apparent surface brightness is dimmer
than \bhal\ with redshift by a factor of $(1+z)^{-4}$.  The distribution of \bhal\ is shown on (Figure~\ref{fig:properties}e), and the formal uncertainty for \bhal\ is about 0.05 dex.
We will use this definition of \bhal\ to compare with a toy model in next section.
We use the axial ratio in the
exponential fit of the galaxy ($AB_{exp}$ given by SDSS
pipeline) as a surrogate for the disc inclination (Figure~\ref{fig:properties}d).
While the axial ratio $b/a$  close to 1 means that the disc galaxy is face-on,
$b/a$ close to 0 means the disc is edge-on.  The typical error
of $b/a$ is about 0.02.  Note that Petrosian radius is defined as the radius
at which the ratio (equals to some specified value,
0.2 for SDSS) of the averaged surface-brightness in a local annulus to the mean
surface-brightness within it (Blanton et al. 2001; Yasuda et al. 2001).
Thus the Petrosian radius is not very sensitive to the disc inclination.
However, the \hal\ surface brightness \bhal\ estimated in Equation~\ref{eq:br}
is only a crude average approximation, because \hii\ regions likely
occupy only a portion of the galaxy, and the distribution may be
very inhomogeneous.

The stellar mass $M_*$ (Figure~\ref{fig:properties}f) is calculated
from template fits to the SDSS five-band photometry with the \texttt{sdss\_kcorrect}
package (Blanton \& Roweis 2007) in IDL.  The magnitudes are corrected to
the AB system (-0.036, 0.012, 0.010, 0.028 and 0.040 for the \textrm{ugriz} filters, respectively)
and corrected for Galactic extinction.  In short, Blanton \& Roweis (2007)
built their five templates using a technique of nonnegative matrix factorization
based on a set of 450 instantaneous burst stellar population models (Bruzual \& Charlot 2003) and
35 emission templates of MAPPINGS-III models (Kewley et al. 2001).  Specifically,
the stellar population models include all six metallicities ($Z$ from 0.0001 to 0.05)
and 25 ages (from 1 Myr to 13.75 Gyr) based on Chabrier (2003) stellar initial mass
function and the Padova 1994 isochrones.  For each stellar population model,
three different dust models are assumed: no dust extinction, $\tau_V=3$ with
Milky Way-like extinction, or $\tau_V=3$ with Small Maggellanic Cloud-like
extinction.  They assumed a homogeneous dust distribution and shell geometry
for the latter two dust models.  The authors compared their measurements
of stellar masses for SDSS galaxies with those obtained by Kauffmann et al. (2003a), and
found that the two set of masses are consistent with each other, with a scatter of only
0.1 dex (refer to their Figure 17 in Blanton \& Roweis 2007).

In addition to these parameters, we also look into other properties based on
the spectra.  The 4000\ \AA\ discontinuity is due to the opacity from ionized metals,
and its amplitude indicates the stellar population ages.  In hot stars,
the metal elements are multiply ionized and the opacity decreases, thus
the 4000\ \AA\ break strength is small for young stellar populations and
large for old, metal-rich populations (Bruzual 1983; Balogh et al. 1999).
The break discontinuity is generally defined as the ratio of the continuum
red- and blue-wards 4000\ \AA.  We adopt the narrow definition D$_n$(4000)
introduced by Balogh et al. (1999) using bands 3850-3950\ \AA\ and
4000-4100\ \AA. D$_n$(4000) is measured from the reddening-corrected modeled
spectrum, which is obtained in the continuum fits described in
Section~\ref{sec:specfit}, instead of directly from the observed spectrum.  An
extensive test shows that the former has two advantages over the latter.
One is that for the observed spectrum of a low S/N ratio, the model-fitting
provides a filter to the noises on the observed spectrum.  The other is,
on the modeled spectrum, the bias in D$_n$(4000) introduced by reddening can
be corrected, about an offset of 0.03 larger for the typical reddening
($\ebv=0.23$) of our sample.  H$\delta$ absorption line equivalent width (EW),
which reaches a peak for the stellar population of age $\sim0.1-1$ Gyr (the
Lick index H$\delta_A$, Worthey \& Ottaviani 1997; Kauffmann et al. 2003a), is
measured from the modeled continuum as well.  The electron density is estimated
from the ratio of \sii $\lambda\lambda 6716, 6731$ \AA\AA\ (Osterbrock  \& Ferland 2006).

\section{RELATION BETWEEN INTRINSIC REDDENING AND OTHER GALAXY PROPERTIES}
\label{sec:relation}

In this section, we will investigate the correlations between the
intrinsic dust-reddening and other observable properties or those
properties that could be deduced from observations.  The properties
include the gas metallicity, the attenuation-corrected \hal\
luminosity or surface brightness, stellar mass, and the inclination of the disc.

For the $\sim$22 000 star-forming disc-dominated galaxies in our final sample,
the distributions of emission-line reddening, which is parameterized
by \ebv, are shown as a function of \lhal\ (Figure~\ref{fig:ebv-lha-z}a)
or \zoh\ (Figure~\ref{fig:ebv-lha-z}b).  On Figure~\ref{fig:ebv-lha-z}b, the
metallicity estimated from $N2$ has a cut-off at $\zoh=9.16$, which corresponds
to $N2=-0.4$.  The histogram of \ebv\ distribution is shown on
Figure~\ref{fig:ebv-lha-z}(c).  While the \ebv\ range is wide
(0.0$\sim$1.0 mag), 95 percent of our sample is in the range of 0.05 to
0.6 mag.

\begin{figure}
\includegraphics[width=0.5\textwidth]{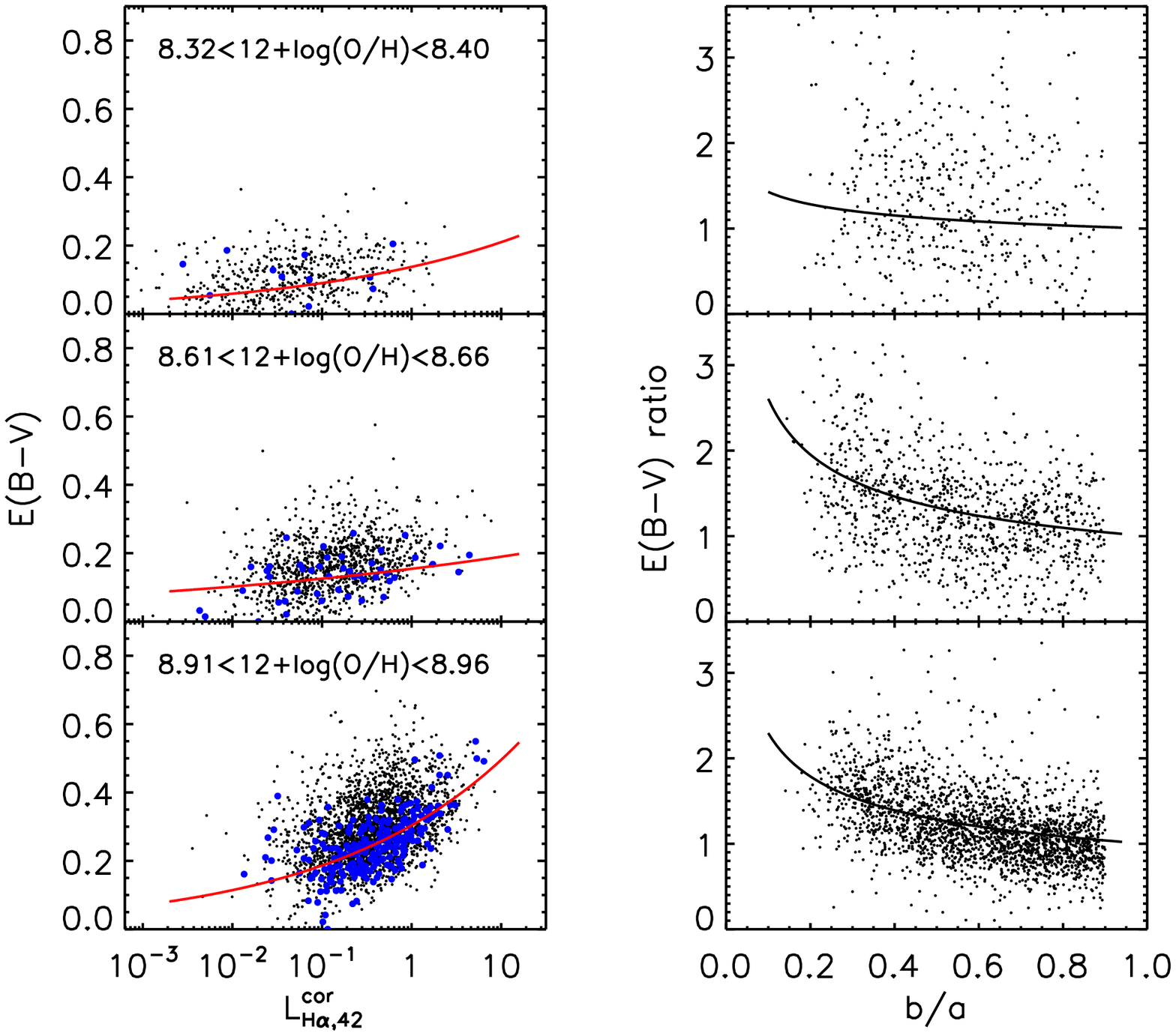}
\caption{Left panel: $\log\ \lhal$ versus \ebv. \lhal\ is in units of \ergs.
Black dots represent for the galaxies with
axial ratio $b/a<0.9$, and blue dots for the face-on galaxies
($b/a>0.9$).  Red lines are the best-fitting power-law for face-on
galaxies in each metallicity bin.  Right panel: The \ebv\ ratio
versus the axial ratio {\em b/a} for each metallicity bin.  The
\ebv\ ratio is the ratio of \ebv\ in other galaxies to
that deduced from the linear function of \lhal\ for face-on
galaxies.  The solid line presents the best-fitting.  The three metallicity
bins are 0.5\zsun, \zsun\ and 2\zsun, as marked on the left panel. }
\label{fig:lhaebvzbin}
\end{figure}

Figure~\ref{fig:ebv-lha-z} shows that the reddening increases
both with increasing \lhal\ and \zoh.  The intrinsic dust-reddening
is clearly correlated with the \hal\ luminosity, with a Spearman rank
coefficient $r_{s}=0.59$ (the probability of null hypothesis
$P_{\rm null}<10^{-5}$) for our sample of star-forming disc galaxies
(Figure~\ref{fig:ebv-lha-z}a).  This relation has been known for
a long time (e.g. Wang \& Heckman 1996; Calzetti et al. 2001;
Hopkins et al. 2001; Afonso et al. 2003).  The reddening correlates with
\hal\ surface brightness as well (not shown in the figure),
with $r_{s}=0.51$ ($P_{\rm null}<10^{-5}$).  On the other hand,
the reddening is also correlated with the gas-phase metallicity
indicated by oxygen abundance \zoh\ ($r_{s}=0.67$, $P_{\rm null}<10^{-5}$,
Figure~\ref{fig:ebv-lha-z}b), which has also been known previously (e.g.
Heckman et al. 1998; Boissier et al. 2004; Asari et al. 2007).  At low
metallicity, the reddening increases slowly with [O/H], but at
high metallicity, it rises more sharply.  The correlation between
\ebv\ and metallcity is the strongest among all of the correlations,
therefore we divide our sample into bins of metallicity to investigate
the relations of reddening with other parameters (\hal\ luminosity/surface
brightness, axial ratio, stellar mass, etc.) at a given metallicity
bin in the rest part of this section.

We divide the sample into continuous metallicity bins.  The bin is so chosen
that a bin-size is no less than 0.05 in \zoh, with at least 400 objects
in each bin.  In order to isolate the inclination-dependence, we check
the relation between \ebv\ and \lhal\ for nearly face-on galaxies, i.e.
$(b/a)>0.90$, and then compare it with the rest of the galaxies.  Clearly,
face-on galaxies have systematically small \ebv\ values than the others at
a given intrinsic \lhal\ (Figure~\ref{fig:lhaebvzbin}, left panel).
We fit the relation of reddening with luminosity for these face-on galaxies
using a power-law function.  For a galaxy, the \ebv\ ratio is defined as the
ratio of \ebv\ of this galaxy, to the expected \ebv\ based on its \hal\
luminosity from the best-fitting power-law function for the face-on galaxies.
The \ebv\ ratio is found to correlate with the axial ratio, which can be well
fitted by $(b/a)^\gamma$.  Figure~\ref{fig:lhaebvzbin} shows $\log \lhal$
versus \ebv\ as well as and the best-fitting curve for the face-on objects
on the left panel, and the dependency of the \ebv\ ratio on the axial ratio
{\em b/a} on the right panel.  These results are only illustrated for three
metallicity bins: $\zoh \simeq 8.36$, 8.64 and 8.94 (i.e. $\sim$0.5\zsun, \zsun\ and
2\zsun, adopting $[\zoh]_{\odot}=8.66\pm0.05$, Asplund et al. 2004).
The correlation between the \ebv\ ratio and the axial ratio becomes
stronger as the metallcity increases, for example $r_s=-0.49$ ($P_{\rm null}<10^{-5}$)
for the bottom metallicity bin $\zoh \simeq 8.94$.
We fit the data in each bin with a joint function in the form of
\begin{equation}
\ebv\ = p L_{\mathrm{H\alpha},42}^q (b/a)^\gamma ,
\label{eq:powfit}
\end{equation}
where $L_{\mathrm{H{\alpha}},42} \equiv \lhal/(10^{42}~ {\rm \ergs})$.  For each
metallicity bin, the fitting parameters are obtained by minimizing $\chi^2$ in the
fitting implemented by \texttt{MPFIT}, and the results are listed in Table~\ref{tab:lha}.
Only uncertainties in \ebv\ have been considered in the fitting, thus the errors of the
fitting parameters are underestimated.

\begin{figure*}
\includegraphics[width=0.3\textwidth]{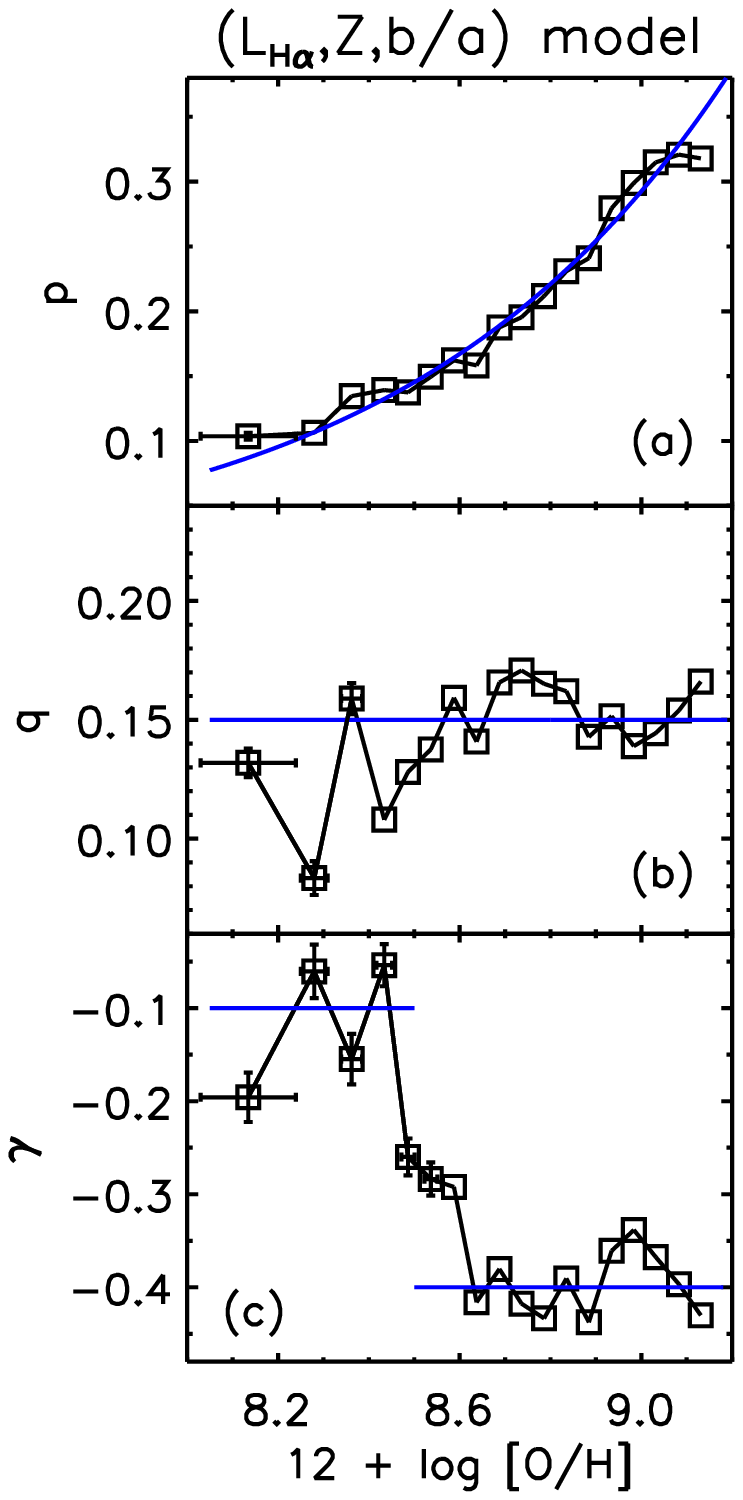}
\includegraphics[width=0.3\textwidth]{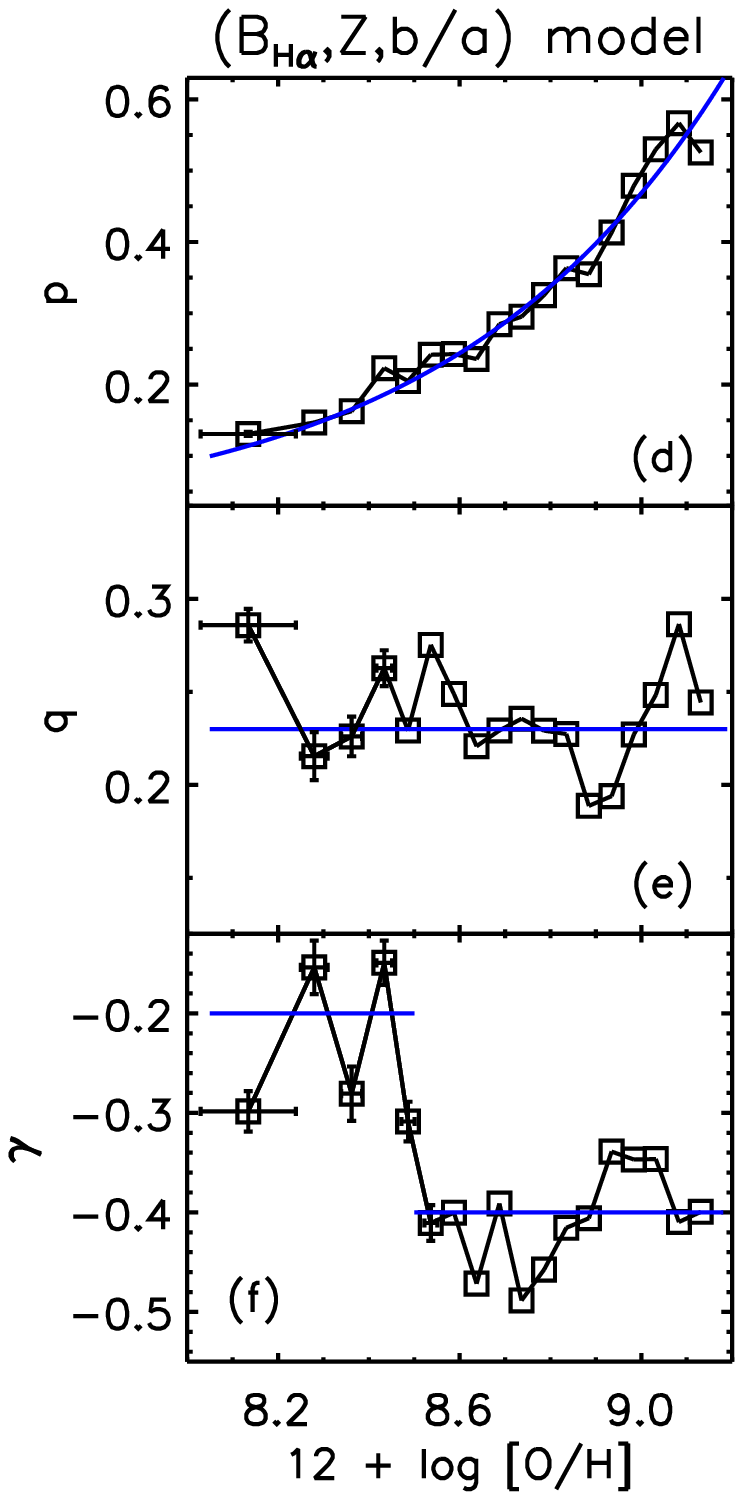}
\caption{Fitting parameters as a function of metallicity. (a) $p$, (b) $q$ and (c) $\gamma$ are
from  Equation~\ref{eq:powfit}, and (d) $p$, (e) $q$ and (f) $\gamma$ are from Equation~\ref{eq:powfit_br}.
Those symbols without error bars indicates the error is smaller than the symbol size. In
each panel, the best-fitting function is shown in blue
solid line.} \label{fig:fitpar-z}
\end{figure*}

\begin{figure*}
\includegraphics[width=0.7\textwidth]{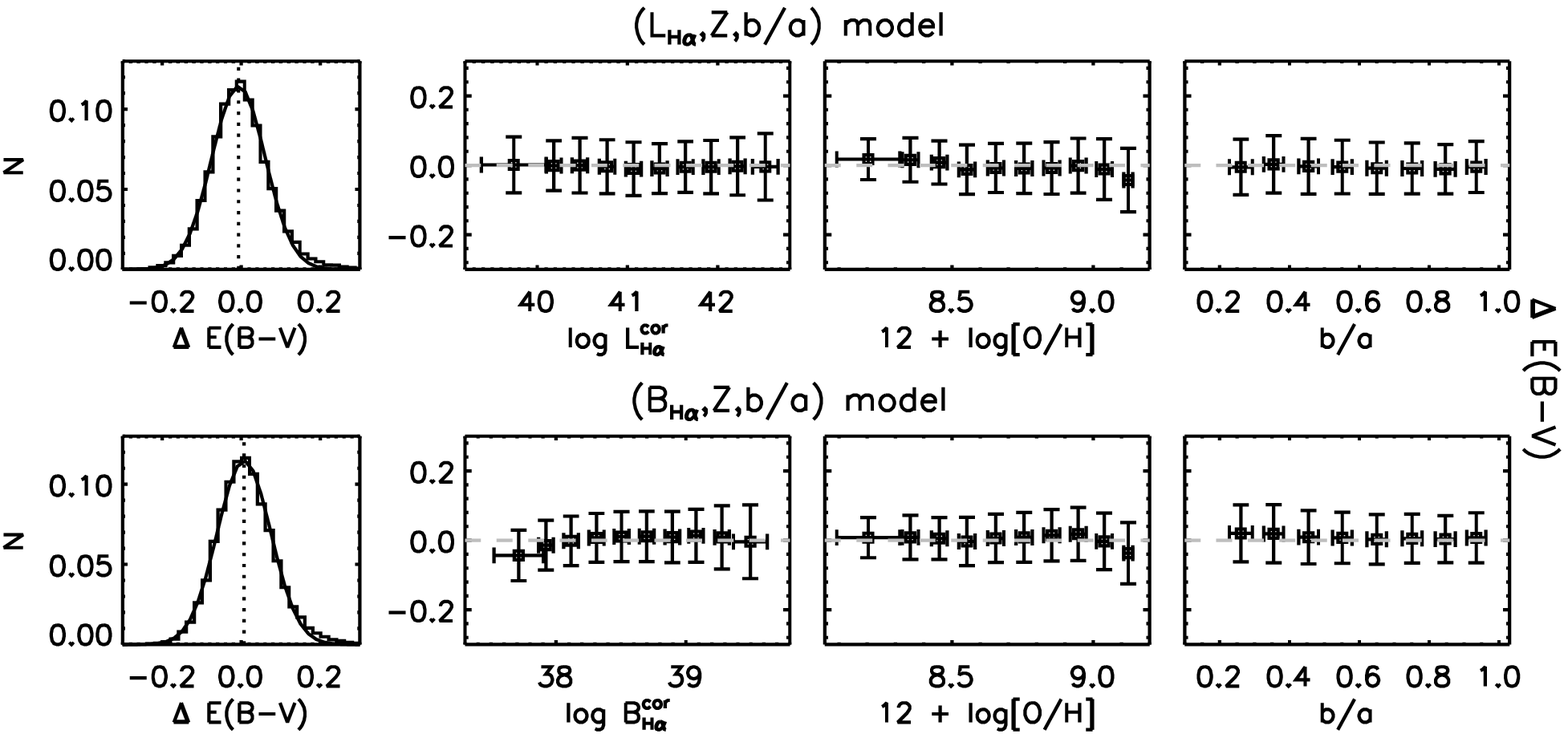}
\caption{The first column shows the distribution of \ebv\ residuals in
our sample for the two empirical models: the (\lhal, $Z$, {\em b/a}) model
and the (\bhal, $Z$, {\em b/a}) model.  \debv\ is defined as
$\debv \equiv \ebv_{obs} - \ebv_{model}$. $N$ is the fraction of the total sample.
The best-fitting Gaussian is over-plotted with the centre shown in dotted
line for each model.  The second to forth column shows \debv\ as a function
of \lhal\ or \bhal\ (in units of log \ergs\ or \ergs kpc\persq), \zoh\ and
the axial ratio $b/a$, respectively.  In each panel, the sample is binned in
corresponding parameters, and the median values with 1$\sigma$ dispersions
in each bin are shown. The gray dashed line represents $\debv=0$.}
\label{fig:debv-dist}
\end{figure*}

Figure~\ref{fig:fitpar-z} shows $p$, $q$ and $\gamma$ as a function of the
metallicity.  We can see that $p$ increases with the metallicity, and can be
fitted with an exponential function of \zoh, i.e. a power-law function of
metallicity $Z$, in the form of
\begin{eqnarray}
p\ =  0.072^{\pm0.001}\exp(\frac{x}{0.71^{\pm0.01}}) \nonumber \\
   =  0.18^{\pm0.02} \left( \frac{Z}{\zsun} \right) ^{0.61^{\pm 0.01}}  ,
\label{eq:plha}
\end{eqnarray}
where $x = \zoh\ - 8$.  The power-law index $q$ varies slightly between 0.08
and 0.17, and the average value is about 0.15.  We therefore adopt the
parameter $q$ as a constant:
\begin{equation}
\langle q\rangle=0.15 .
\end{equation}
$|\gamma|$ first increases from about 0.1 with metallicity and then reaches
to a constant ($|\gamma| \sim 0.4$) at about $\zoh \simeq 8.5$.  We use a step
function to fit $\gamma$ as a function of \zoh, which is expressed as
\setlength\arraycolsep{0.5mm}
\begin{eqnarray}
\gamma\ = & -0.1 \qquad for \; 0.1 < x < 0.5     &   \nonumber \\
        = & -0.4 \qquad for \; 0.5 \leq x < 1.2  &  .
\label{eq:gamma}
\end{eqnarray}
Substitute $p$, $q$ and $\gamma$ into Equation~\ref{eq:powfit}, we will get the
overall function of luminosity, metallicity and inclination, and we call this
(\lhal, $Z$, {\em b/a}) model hereafter.  To verify the reliability of our
empirical formulae, we define the \ebv\ residual as $\debv \equiv \ebv_{obs} - \ebv_{model}$,
and show its distribution in Figure~\ref{fig:debv-dist} (the first column,
upper panel).  The \ebv\ distribution can be well fitted with a Gaussian
profile, with width $\sigma=0.068$ mag.  If we adopt a more stringent S/N
criteria (S/N$>100$) on \hal\ emission, the result will yield a sample of
$\sim$5000 star-forming disc galaxies.  We repeat the procedure above and
find the scatter of \debv\ remains similar, $\sigma=0.066$ mag, indicating
that the scatter is independent of \hal\ S/N, in the S/N range we consider
(we will discuss in more details and show that our S/N criteria is reasonable
in the following Section~\ref{sec:erranaly}).  Besides, we find that \debv\
is uncorrelated with the parameters including \lhal, metallicity or axis
ratio (Figure~\ref{fig:debv-dist}, the right three columns), which suggests
that the (\lhal, $Z$, {\em b/a}) model considered above reproduces the
intrinsic reddening well in the disc galaxies.

Note that at low metallicities the scatter in the relation between \ebv\ ratio
and $b/a$ is larger than that at high metallicities (Figure~\ref{fig:fitpar-z},
the right column).  The power-law index $\gamma$ is close to zero at low metallcity,
i.e. the relationship between \ebv\ ratio and $b/a$ is almost flat; while
$\gamma \sim -0.4$ when the metallicity increases to solar or super-solar
value.  This suggests that dust reddening is not so sensitive to the axis
ratio at low metallicity as at high metallcity.  It may be attributed to two
possible reasons.  One is that the relation at low metallicity might remain
as strong as at high metallcity, but the limited dynamic-range of \ebv\ and
the measurement uncertainties prevent us from recovering the relationship.
The \ebv\ dynamic-range is only about 0.2 mag at low metallicity, and increases
to 0.8 mag at high metallicity.  The other possible reason is that the relation
indeed becomes weaker at low metallicity.  At present we can not distinguish
which one is true, especially when the number of face-on galaxies at low metallicity
is limited.  If we adopt $\gamma$ to be $-0.4$ for all metallicities, the
distribution of \debv\ slightly changes.

\setcounter{table}{1}
\begin{table*}
 \centering
 \begin{minipage}{120mm}
  \caption{Coefficients of empirical fits to reddening in a function as in
  Equation~\ref{eq:powfit}, the empirical relation of reddening parameterized
  by color excess \ebv, as a function of \lhal\ and the axial ratio {\em b/a}
  in different metallicity bins.
}
  \label{tab:lha}
  \begin{tabular}{@{}rrcccc@{}}
  \hline
  Metallcity bin & Num &  $p$ & $q$ & $\gamma$ & $\chi^2_{\nu}$  \\
 \hline
$     \zoh<8.22$ &  423 & $ 0.104^{\pm0.003}$ & $ 0.132^{\pm0.006}$ & $-0.196^{\pm0.027}$ & 10.75 \\
$8.22<\zoh<8.32$ &  448 & $ 0.106^{\pm0.003}$ & $ 0.083^{\pm0.007}$ & $-0.060^{\pm0.029}$ &  9.14 \\
$8.32<\zoh<8.40$ &  515 & $ 0.135^{\pm0.003}$ & $ 0.159^{\pm0.007}$ & $-0.155^{\pm0.027}$ &  7.01 \\
$8.40<\zoh<8.46$ &  539 & $ 0.139^{\pm0.003}$ & $ 0.108^{\pm0.006}$ & $-0.054^{\pm0.023}$ &  7.41 \\
$8.46<\zoh<8.51$ &  568 & $ 0.138^{\pm0.002}$ & $ 0.128^{\pm0.005}$ & $-0.260^{\pm0.020}$ &  5.84 \\
$8.51<\zoh<8.56$ &  733 & $ 0.150^{\pm0.002}$ & $ 0.138^{\pm0.005}$ & $-0.284^{\pm0.018}$ &  6.65 \\
$8.56<\zoh<8.61$ &  819 & $ 0.162^{\pm0.002}$ & $ 0.159^{\pm0.004}$ & $-0.292^{\pm0.014}$ &  7.26 \\
$8.61<\zoh<8.66$ & 1088 & $ 0.158^{\pm0.002}$ & $ 0.141^{\pm0.004}$ & $-0.416^{\pm0.012}$ &  6.20 \\
$8.66<\zoh<8.71$ & 1329 & $ 0.188^{\pm0.002}$ & $ 0.166^{\pm0.003}$ & $-0.380^{\pm0.010}$ &  6.51 \\
$8.71<\zoh<8.76$ & 1583 & $ 0.195^{\pm0.001}$ & $ 0.171^{\pm0.003}$ & $-0.418^{\pm0.009}$ &  6.09 \\
$8.76<\zoh<8.81$ & 1917 & $ 0.212^{\pm0.001}$ & $ 0.165^{\pm0.003}$ & $-0.433^{\pm0.008}$ &  7.09 \\
$8.81<\zoh<8.86$ & 2162 & $ 0.231^{\pm0.001}$ & $ 0.162^{\pm0.002}$ & $-0.391^{\pm0.006}$ &  6.97 \\
$8.86<\zoh<8.91$ & 2456 & $ 0.241^{\pm0.001}$ & $ 0.143^{\pm0.002}$ & $-0.438^{\pm0.006}$ &  7.02 \\
$8.91<\zoh<8.96$ & 2551 & $ 0.279^{\pm0.001}$ & $ 0.152^{\pm0.002}$ & $-0.361^{\pm0.005}$ &  8.27 \\
$8.96<\zoh<9.01$ & 2191 & $ 0.299^{\pm0.001}$ & $ 0.139^{\pm0.002}$ & $-0.338^{\pm0.005}$ &  8.97 \\
$9.01<\zoh<9.06$ & 1652 & $ 0.315^{\pm0.001}$ & $ 0.144^{\pm0.002}$ & $-0.368^{\pm0.006}$ & 10.64 \\
$9.06<\zoh<9.11$ & 1052 & $ 0.321^{\pm0.001}$ & $ 0.154^{\pm0.002}$ & $-0.397^{\pm0.007}$ & 13.53 \\
$     \zoh>9.11$ &  590 & $ 0.318^{\pm0.002}$ & $ 0.166^{\pm0.003}$ & $-0.430^{\pm0.009}$ & 11.33 \\
\hline
\end{tabular}
\end{minipage}
\end{table*}

\begin{figure}
\includegraphics[width=0.5\textwidth]{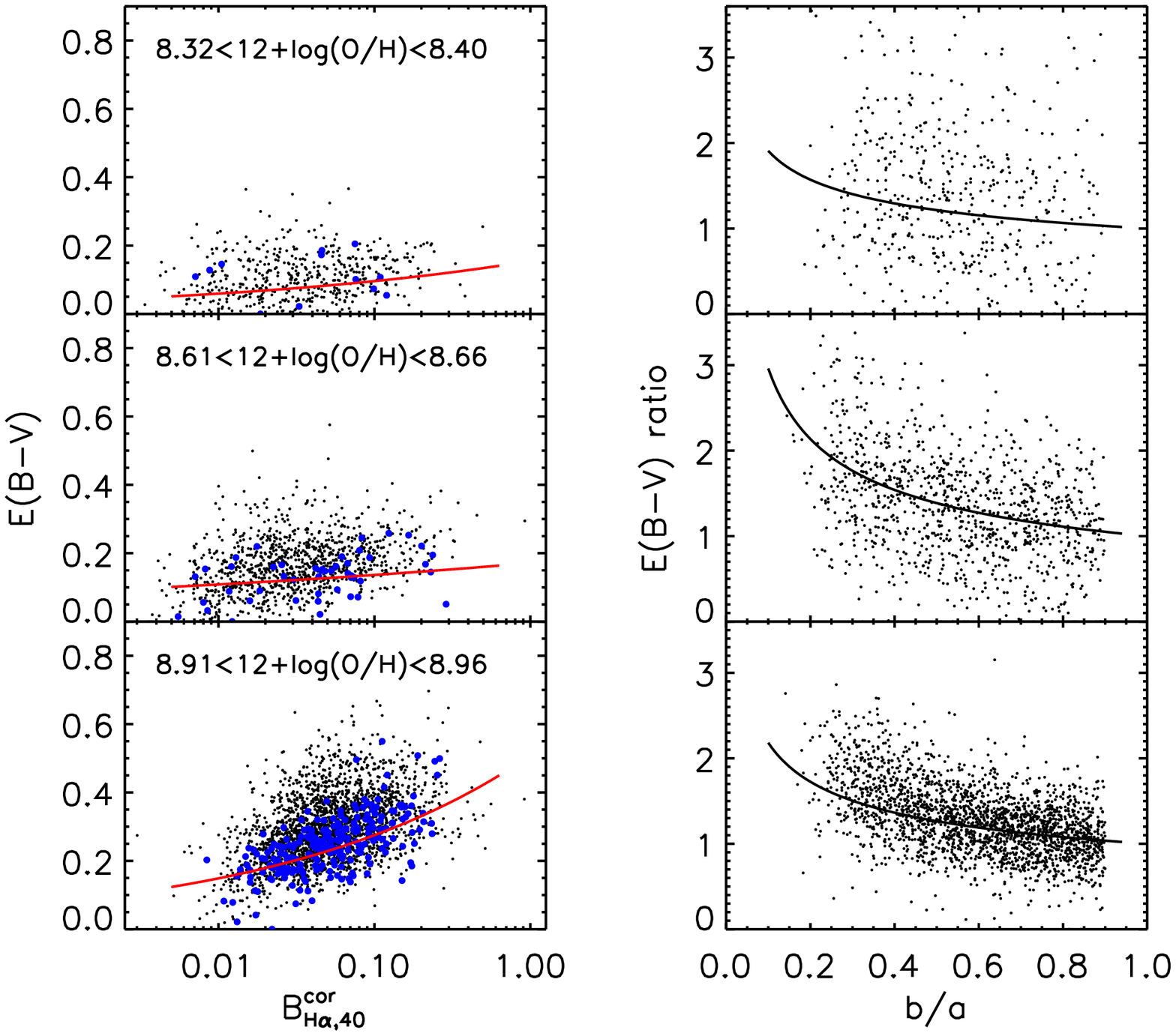}
\caption{Left panel: $\log\ \bhal$ versus \ebv.  \bhal\ is in units of
\ergs\ kpc \persq.  Right panel: The \ebv\ ratio versus the axial ratio
{\em b/a} for each metallicity bin.  The \ebv\ ratio is the ratio of
\ebv\ in other galaxies to that deduced from the linear function of
\bhal\ for face-on galaxies.  The three metallicity bins are 0.5\zsun,
\zsun\ and 2\zsun, as marked on the left panel.  The symbols and the
lines represent the same as in Figure~\ref{fig:lhaebvzbin}.
}
\label{fig:brebvzbin}
\end{figure}

We do similar analysis on how \ebv\ varies as a function of \hal\ surface
brightness and inclination in the metallicity bins, because the
\hal\ (SFR) surface density is believed to physically link with the gas
surface density, which is closely coupled with dust.  Besides, the gas
surface density is provided by the current models of galaxy formation.
Therefore, the \ebv\ as a function of \hal\ surface brightness
will enable direct application of the empirical relation on the intrinsic
reddening in the galaxy evolution models.  Figure~\ref{fig:brebvzbin}
shows \bhal\ versus \ebv\ and the best-fitting curve of a power-law function
for the face-on objects on the left panel, while the distribution of \ebv\
ratio against the axial ratio for the three selected metallicity bins on the
right panel.  In each metallcity bin, we fit the data with a function in the
form of
\begin{equation}
\ebv\ = p B_{\mathrm{H\alpha},40}^q (b/a)^\gamma ,
\label{eq:powfit_br}
\end{equation}
where $B_{\mathrm{H{\alpha}},40} \equiv \bhal/(10^{40}~ {\rm \ergs kpc\persq)}$.
All the fitting parameters are listed in Table~\ref{tab:br}.  We also fit $p$
with an exponential function of \zoh, $q$ with a constant, and
$\gamma$ as a step function (Figure~\ref{fig:fitpar-z}).
\begin{eqnarray}
p\ = 0.092^{\pm0.001}\exp(\frac{x}{0.61^{\pm0.01}}) \nonumber  \\
   = 0.27^{\pm0.02} \left( \frac{Z}{\zsun} \right) ^{0.71^{\pm0.01}}  ,
\label{eq:pbr}
\end{eqnarray}
\begin{equation}
\langle q\rangle=0.23 ,
\end{equation}
and
\setlength\arraycolsep{0.5mm}
\begin{eqnarray}
\gamma\ = & -0.2 \qquad for \; 0.1 < x < 0.5     &   \nonumber \\
        = & -0.4 \qquad for \; 0.5 \leq x < 1.2  &   .
\label{eq:gamma_br}
\end{eqnarray}
With the model expressed in Equation~\ref{eq:powfit_br}, hereafter (\bhal,
$Z$, {\em b/a}) model, we calculate \debv\ and show the distribution in
Figure~\ref{fig:debv-dist}.  The distribution can be also fitted well with
a Gaussian profile, with width $\sigma=0.068$~mag.  \debv\ is uncorrelated
with the parameters including \bhal, the metallicity or the axial ratio,
which suggests that the (\bhal, $Z$, {\em b/a}) model can reproduce the
intrinsic reddening well in the disc galaxies too.

\setcounter{table}{2}
\begin{table*}
 \centering
 \begin{minipage}{120mm}
  \caption{Coefficients of empirical fits to reddening in a function as in
  Equation~\ref{eq:powfit_br}, the empirical relation of \ebv\ as a function of
  \bhal\ and the axial ratio {\em b/a} in different metallicity bins.
}
  \label{tab:br}
  \begin{tabular}{@{}rrcccc@{}}
  \hline
  Metallcity bin & Num & $p$ & $q$ & $\gamma$ & $\chi^2_{\nu}$  \\
 \hline
$     \zoh<8.22$ &  423 & $ 0.131^{\pm0.003}$ & $ 0.286^{\pm0.009}$ & $-0.298^{\pm0.020}$ &  9.85 \\
$8.22<\zoh<8.32$ &  448 & $ 0.147^{\pm0.006}$ & $ 0.215^{\pm0.013}$ & $-0.154^{\pm0.027}$ &  8.78 \\
$8.32<\zoh<8.40$ &  515 & $ 0.163^{\pm0.005}$ & $ 0.226^{\pm0.011}$ & $-0.281^{\pm0.027}$ &  7.24 \\
$8.40<\zoh<8.46$ &  539 & $ 0.223^{\pm0.006}$ & $ 0.263^{\pm0.010}$ & $-0.149^{\pm0.022}$ &  6.54 \\
$8.46<\zoh<8.51$ &  568 & $ 0.205^{\pm0.005}$ & $ 0.229^{\pm0.008}$ & $-0.309^{\pm0.020}$ &  5.34 \\
$8.51<\zoh<8.56$ &  733 & $ 0.242^{\pm0.006}$ & $ 0.275^{\pm0.008}$ & $-0.411^{\pm0.018}$ &  5.97 \\
$8.56<\zoh<8.61$ &  819 & $ 0.243^{\pm0.004}$ & $ 0.249^{\pm0.006}$ & $-0.400^{\pm0.015}$ &  6.91 \\
$8.61<\zoh<8.66$ & 1088 & $ 0.236^{\pm0.004}$ & $ 0.221^{\pm0.005}$ & $-0.472^{\pm0.012}$ &  5.89 \\
$8.66<\zoh<8.71$ & 1329 & $ 0.285^{\pm0.004}$ & $ 0.229^{\pm0.004}$ & $-0.391^{\pm0.010}$ &  6.31 \\
$8.71<\zoh<8.76$ & 1583 & $ 0.295^{\pm0.004}$ & $ 0.236^{\pm0.004}$ & $-0.488^{\pm0.010}$ &  5.93 \\
$8.76<\zoh<8.81$ & 1917 & $ 0.325^{\pm0.003}$ & $ 0.229^{\pm0.003}$ & $-0.457^{\pm0.008}$ &  6.43 \\
$8.81<\zoh<8.86$ & 2162 & $ 0.363^{\pm0.003}$ & $ 0.227^{\pm0.003}$ & $-0.415^{\pm0.007}$ &  6.58 \\
$8.86<\zoh<8.91$ & 2456 & $ 0.355^{\pm0.003}$ & $ 0.189^{\pm0.003}$ & $-0.406^{\pm0.006}$ &  6.77 \\
$8.91<\zoh<8.96$ & 2551 & $ 0.413^{\pm0.003}$ & $ 0.194^{\pm0.002}$ & $-0.339^{\pm0.005}$ &  8.24 \\
$8.96<\zoh<9.01$ & 2191 & $ 0.479^{\pm0.003}$ & $ 0.227^{\pm0.003}$ & $-0.347^{\pm0.005}$ &  8.42 \\
$9.01<\zoh<9.06$ & 1652 & $ 0.530^{\pm0.003}$ & $ 0.248^{\pm0.003}$ & $-0.346^{\pm0.006}$ &  9.26 \\
$9.06<\zoh<9.11$ & 1052 & $ 0.567^{\pm0.004}$ & $ 0.286^{\pm0.003}$ & $-0.410^{\pm0.007}$ & 10.98 \\
$     \zoh>9.11$ &  590 & $ 0.526^{\pm0.005}$ & $ 0.244^{\pm0.004}$ & $-0.399^{\pm0.009}$ & 11.22 \\
\hline
\end{tabular}
\end{minipage}
\end{table*}

In order to investigate the origin of the scatter in the empirical relations,
we examine the correlation between \debv\ and other galaxy parameters, such
as stellar mass, the 4000\ \AA\ break index, EW of Balmer
absorption lines, the electron density in \hii\ region or concentration
index.  The \debv\ for either (\lhal, $Z$, {\em b/a}) model or
(\bhal, $Z$, {\em b/a}) model is not correlated with any of the parameters
above, implying that the dust reddening in a disc galaxy is primarily
determined by only three physical properties: luminosity (SFR) / surface
brightness (SFR surface density), metallicity and disc inclination.

We also attempt another multi-parameter model
($M_*$, $Z$, {\em b/a}) with similar procedure described above, and find
that the scatter of \debv\ is 0.072 mag, which is slightly larger than
the two models above (Equation~\ref{eq:powfit} and \ref{eq:powfit_br}).
However, \debv\ for this model is still weakly correlated with \lhal\ ($r_s=0.15$,
$P_{\rm null}<10^{-5}$).  This suggests that (\lhal\ or SFR, $Z$) combination
works better than ($M_*$, $Z$) combination.  Physically, at a given metallicity,
the amount of dust is proportional to the amount of cold gas, which is closely
correlated with SFR.

\subsection{Comparison with previous work}

We compare our models with previous empirical relations.  Garn \& Best (2010)
investigate the dependency of dust extinction at \hal\ on SFR, metallicity
and stellar mass respectively, in star-forming galaxies based on SDSS database.
They conclude that the stellar mass is the most fundamental parameter
in determining the dust extinction in the local Universe, and their relation
between stellar mass and dust extinction seems to hold up to redshift of 1.5 (Sobral et al. 2011).
They utilize different models to fit the extinction \ahal, with each model
taking into consideration one of the three parameters: SFR, $M_*$ and metallicity.
They find that the $M_*$ model gives the best fit, with the minimum of
scatter (0.28 mag) in the extinction residual
$\Delta\ahal \equiv \ahal^{obs} - \ahal^{model}$, less than the
scatter for SFR model and metallicity model, both of which are 0.33 mag.
When calculating the \ahal, they adopt the Calzetti et al. (2000) dust
attenuation-curve. We repeat their analysis for our selected sample of disc
galaxies and compare the scatter of $\Delta\ahal$ among different models.  In each
model a low-order (third order) polynomial is adopted to fit the extinction,
and the distribution of $\Delta\ahal$ is well fitted by a Gaussian, with widths
$\sigma=$0.27, 0.28 and 0.28 mag for the $M_*$, \lhal\ (SFR) and metallicity
model respectively.  Higher order polynomial fits will not decrease the scatter.
We check the scatter of $\Delta\ahal$ for our (\lhal, $Z$, {\em b/a}) or
(\bhal, $Z$, {\em b/a}) model assuming Calzetti's attenuation law, and the
Gaussian width is only $\sigma=0.22$ mag (corresponding to $\sigma=0.065$ mag
for the \debv\ distribution).  We also check for the ($M_*$, $Z$, {\em b/a}) model,
and find that the scatter of $\Delta\ahal$ is 0.24 mag, which is better than the
single-parameter models, but worse than the other two multi-parameter models.

Our empirical relations predict dust reddening better than the single-parameter
models, probably because we include disc inclination in the analysis.  In order to
investigate what dominates the relationships of the variables, we apply principle
component analysis on the physical variables.  The derived principle components (PCs) are
linear combinations of the variables, and represent for the directions of maximum
variance in the data.  For \hal\ luminosity and surface brightness,
\ebv, $M_*$ and metallicity, the quantities are normalized to zero mean and standard
deviation of unity, while for the axial ratio original value is used.  We test with
different subsets of variables adopted in each multi-parameter model as well as all of
the variables, and find similar results.  Take the case of (\lhal, $Z$, {\em b/a}) model
for example, the first two PCs contribute to about 80 percent of the total variance.
In PC1, the variables \lhal, $Z$ and \ebv\ weight almost equivalently.  This is consistent
with what Garn \& Best (2010) found.  But in PC2 the axial ratio weights the most (-0.98) and
\ebv\ weights the second (0.30).  The result is consistent with what we previously found.
Although the axial ratio only weakly correlated with \ebv\ ($r_s=-0.13$, $P_{\rm null}<10^{-5}$),
after removing the \ebv\ dependence on \hal\ luminosity and metallicity, the
negative correlation between $b/a$ and \ebv\ emerges.

\subsection{Error analysis}
\label{sec:erranaly}

We have shown that the global reddening of the emission lines from a disc
galaxy can be reasonably determined by its \hal\ luminosity or surface
brightness, gas metallicity and disc inclination.  The $1\sigma$ scatter
of \debv\ is about 0.07 mag, which is contributed by several sources, including
the measurement uncertainty in the Balmer decrement, the error in the
metallicity, the uncertainties in other variables and the intrinsic
scatter in the relation.  We will discuss the uncertainties in the following.

Let us take a review of the uncertainties in the physical quantities.
The formal error for \lhal\ measurement is 0.04 dex, and for \bhal\ is 0.05 dex.
As discussed in Section~\ref{sec:lhacor}, the uncertainty induced by
aperture-correction and attenuation-correction is less than 0.12 dex.  The
typical error of $b/a$ is 0.02.  The uncertainty in metallicity calibration dominates the
uncertainty of metallicity, estimated to be $\sim 0.15$ dex.  We use Monte Carlo
simulation to estimate the error of \ebv\ for Equation~\ref{eq:powfit} and
\ref{eq:powfit_br} taking the uncertainties for the best-fit
parameters and the measurement errors for the physical variables.  The errors
of \ebv\ predicted by Equation~\ref{eq:powfit} and \ref{eq:powfit_br} are
both about 0.056 mag.  If taking the uncertainty in aperture-correction and
attenuation-correction (0.12 dex at most) into account in \lhal\ and \bhal\ error, the simulated
\ebv\ errors slightly changes to 0.058 mag.  Remember that the measurement error of \ebv\
is 0.03 mag, the scatter of \debv\ (0.07 mag) can be almost explained by the
uncertainties of the physical quantities.
Considering all these uncertainties, \ebv\ predicted from the empirical
relations are quite consistent with the observed values.

From Figure~\ref{fig:debv-dist} we can see that the \debv\ distribution
can be well described by a Gaussian profile, indicative of a random
distribution.  However, the intrinsic scatter of the empirical relation
is still unclear, because of the large uncertainty in the metallicity
calibration.  Therefore, to estimate the intrinsic scatter of the relation,
it will be of help to decrease the uncertainty in the metallicity calibration
and obtain the metallicity more precisely.  One may find there is slight
excess over the positive tail of the gaussian distribution.  We examine
the objects (about one percent of the whole sample) with $\debv>0.2$,
i.e. the observed \ebv\ is under-estimated with our empirical models by 0.2
mag, and find that these objects are prone to large Balmer decrement,
low \hbe\ S/N and low \hbe\ EW.

As previously mentioned, we require S/N$>20$ for \hal, and S/N$>10$
for \hbe\ and \oiii\ when selecting our sample in Section~\ref{sec:sample}.
The S/N criteria for \hal\ will exclude some galaxies with low SFR
and low \hal\ EW.  Thus the number of galaxies with low \hal\ luminosity
and low SFR will be reduced.  In fact, when the criteria S/N$>10$ is applied
for \hbe, most of the remaining sample ($\sim$99.9 percent) will have \hal\
S/N$>20$.  In order to investigate whether the \hbe\ and \oiii\ S/N criteria
bias our results, we attempt a looser S/N($>5$) criteria for both lines, and
the sample is expanded by 50 percent to have 34815 galaxies.  Then we fit the
data with both the (\lhal, $Z$, {\em b/a}) model and (\bhal, $Z$, {\em b/a})
model.  The results show that the fit parameters are quite similar as
those we have obtained in Section~\ref{sec:relation}, but the scatter in
\debv\ increases to about 0.08 mag.  Note that 10 percent uncertainty in
\hal/\hbe\ will result in an uncertainty of 0.086 mag in \ebv, which explains
the increased scatter.  Thus including the low S/N objects merely introduces more
scatter in the relation.

\section{A TOY MODEL ON DUST EXTINCTION TO \hii\ REGIONS IN A DISC GALAXY}
\label{sec:toymodel}

Two different sources contribute to the extinction for \hii\ regions:
dust associated with the individual \hii\ region and diffuse foreground
dust in the galaxy. Let us first consider the dust in the precursor
molecular clouds (MC). The intrinsic extinction in individual \hii\
region depends on the relative distribution of young massive stars, gas and
dust in the star-formation regions, which are also closely related to
the age and size of the precursor MC, as well as the stochastic
formation of high mass stars for a given initial mass function. An
\hii\ region is embedded in the cold gas/dust envelope before the
cold gas/dust is blown away.  The time scale for such a process is
expected to depend on the size and density of the precursor cloud:
longer for more massive and dense \hii\ region.  Panuzzo et al.
(2003) discussed the attenuation to the \hal\ emission line for different
evaporating time scale, in comparison with the lifetime of massive
stars that produce the ionizing continuum.  They simply adopt a central
ionizing star in a spherical cloud, which corresponds to the maximum
attenuation case. If the time scale is short enough, most \hii\
regions are produced outside of the MC, and the main attenuation
is due to foreground dust. In the case of long evaporation time,
an \hii\ region is buried in the MC during most of their life time
and the extinction associated with the \hii\ region will be important.
In a galaxy, all these cases are likely present.  Furthermore, the young
ionizing stars are not necessarily forming in the cloud centre.  These
complications make any prediction of the reddening in an individual cloud
rather difficult.

However, as far as a large number of \hii\ regions are concerned, the
statistical behavior may be quite well defined. The \hal\ luminosity
function of \hii\ regions in spiral galaxies can be described by double
power-law functions, with a flatter power-law below the break luminosity
of 38.6 (in units of \ergs, logarithmic), and a steeper one above it
(e.g. Kennicutt et al. 1989; Bradley et al. 2006).  In our sample, most
galaxies have \hal\ luminosities above 40 (in units of \ergs, logarithmic)
with a median of 41.5, and the maximum is 43.3. Therefore, a galaxy consists
of numerous (order of hundreds to millions of) \hii\ regions.   The \hii\
regions follow certain distributions of ages and initial masses, determined
by the star-formation processes, because the star-formation time in a galaxy
is likely much longer than the lifetime of an individual \hii\ region, i.e.
the life time of OB stars.  If these distributions are the same for each
galaxy and there are a large number of \hii\ regions, one may expect that
the average extinction within \hii\ regions should be similar in each
galaxy. In this respect, it is not expected that the dust reddening
depends on the \hal\ luminosity or the inclination of the disc.  In
contrast, for example, if the MC mass function and stellar IMF depend on the
global properties of the galaxies, such as the gas surface density,
then one may expect that the average extinction is correlated with
the global properties. However, this is poorly known so far.

Next, we consider the extinction due to diffuse interstellar dust in
a galaxy. It is expected that this extinction depends on the gas surface
density, inclination, metallicity of the interstellar medium and size
of the galaxy.  To simplify the model, we will adopt smooth distributions
for the dust, gas and star-formation regions with the following assumptions:
\begin{description}
\item[(1)] dust-to-gas density ratio is proportional to the gas metallicity (Issa
et al. 1990), i.e., gas and dust are coupled:
\begin{equation}
\rho_\mathrm{d}\propto\rho_\mathrm{g}\times Z,
\label{eq:rhod1}
\end{equation}
where $\rho_\mathrm{d}$ and $\rho_\mathrm{g}$ are dust density and gas density,
respectively;
\item[(2)] both gas and dust distributions in the vertical ($z$)
and radial ($R$) direction can be described by an exponential law
(e.g. de Jong 1996; Thomas et al. 2004; Misiriotis et al. 2006), i.e.,
\begin{equation}
\rho_{\mathrm{g}} = \rho_{\mathrm{g}}(0) \exp(-z/h_\mathrm{g}) \exp(-R/R_\mathrm{g}),
\label{eq:rhog}
\end{equation}
where $\rho_{\mathrm{g}}(0)$ is the gas density at $z=0$ and $R=0$, $h_\mathrm{g}$ and
$R_\mathrm{g}$ are the characteristic lengths in the vertical and radial directions for gas distribution.
But the dust distribution may have different length scales due to radial metallicity gradient:
\begin{equation}
\rho_{\mathrm{d}} = \rho_{\mathrm{d}}(0) \exp(-z/h_\mathrm{d}) \exp(-R/R_\mathrm{d}),
\label{eq:rhod2}
\end{equation}
where $\rho_{\mathrm{d}}(0)$ is the dust density at $z=0$ and $R=0$, $h_\mathrm{d}$ and
$R_\mathrm{d}$ are the characteristic lengths in the vertical and radial directions for dust distribution;
\item[(3)] the \hii\ distribution is also exponential in both radial and vertical
direction, but with different characteristic lengths (e.g. de Jong 1996):
\begin{equation}
\rho_\mathrm{\hal} = \rho_\mathrm{\hal}(0) \exp(-z/h_\mathrm{\hal}) \exp(-R/R_\mathrm{\hal}).
\label{eq:rhoha}
\end{equation}
\end{description}
By integrating $\rho_\mathrm{\hal}$ over $z$ direction,
we yield a surface density of \hal, $\Sigma_\mathrm{\hal}$.  Specifically, the \hal\
surface density at $R=0$ is
\begin{eqnarray}
\Sigma_\mathrm{\hal,0}& = & \int_0^\infty \rho_\mathrm{\hal}(0) \exp(-z/h_\mathrm{\hal}) dz +
                           \int_{-\infty}^0 \rho_\mathrm{\hal}(0) \exp(z/h_\mathrm{\hal}) dz \nonumber \\
                      & = & 2\rho_\mathrm{\hal}(0) h_\mathrm{\hal}
\end{eqnarray}

The gas disc is rather thin in general.  As a result, dust attenuation can
be modeled as plane-parallel except at extremely high inclinations.  Young
stars are formed in dense molecular cores, which might be distributed in a plane
thinner than the gas disc.  For a plane-parallel model, the optical
depth is expressed as
\begin{equation}
\tau\ = \rho_{\mathrm{d}} k_\nu z / \cos i,
\end{equation}
where $k_\nu$ is the opacity coefficient, and $i$ is the
inclination angle, and $i=0^{\degr}$ represents the line of sight perpendicular
to the galaxy disc, while $i=90^{\degr}$ represents the line of sight parallel
to the galaxy disc.  The optical depth to \hal\ emission from the middle
plane of the disc to the upper disc surface at $R=0$ is defined as
\begin{equation}
\tau_\mathrm{\hal,0} = \int_0^\infty \rho_{\mathrm{d}}(0) k_\nu \exp(-h/h_\mathrm{d}) dh = \rho_{\mathrm{d}}(0) k_\nu h_\mathrm{d}.
\end{equation}
It is symmetric between the middle plane to the upper and lower disc surface,
thus the overall optical depth through the whole disk
vertically at $R=0$ should be $2\tau_\mathrm{\hal,0}$.  Then we consider the
optical depth from the point at $z$($>0$) on the vertical axis ($R=0$) to the
upper surface of the disc is
\begin{equation}
\tau_{1}(z) = \int_z^\infty \rho_{\mathrm{d}}(0) k_\nu \exp(-h/h_\mathrm{d}) dh = \tau_\mathrm{\hal,0} \exp(-z/h_\mathrm{d}) ,
\label{eq:tau_up}
\end{equation}
and the optical depth from that point to the lower disc surface can be expressed as
\begin{equation}
\tau_{2}(z) = 2\tau_\mathrm{\hal,0} - \tau_{1}(z).
\label{eq:tau_low}
\end{equation}

Light from the \hii\ regions at the position ($z$, $R$) needs to travel through
the disc, from both upper and lower sides.  With these descriptions, we can calculate the
\hal\ luminosity, \hal\ surface brightness and average attenuation at \hal\
and \hbe\ by integrating the radiation transfer function over both vertical
and radial directions within the disc.  Then the \hal\ luminosity is
\begin{eqnarray}
& \lhal\  =  \int_0^2\pi d\theta \int_0^{R_{\hal, max}} RdR \int_0^\infty \rho_\mathrm{\hal} \nonumber  \\
&    \{ \exp[-\frac{\tau_{1}(z)}{\cos i} \exp(-R/R_\mathrm{d})] + exp[-\frac{\tau_{2}(z)}{\cos i} \exp(-R/R_\mathrm{d})] \}.
\label{eq:lhal0}
\end{eqnarray}
Substitue Equation~\ref{eq:rhoha}, \ref{eq:tau_up}, \ref{eq:tau_low},
$y=\exp(-z/h_\mathrm{\hal})$ and $a=R/R_\mathrm{\hal}$ into Equation~\ref{eq:lhal0}, then we get
\begin{eqnarray}
\lhal\ = & \pi R^2_\mathrm{\hal} \Sigma_\mathrm{\hal,0} \int_0^1 dy \int_0^\xi ada \exp(-a) \{ \nonumber \\
         & \exp [-\frac{\tau_\mathrm{\hal,0}}{\cos i}\exp(-a/\eta)y^{1/\kappa}] +   \nonumber \\
         & \hspace*{2em} \exp[-\frac{\tau_\mathrm{\hal,0}}{\cos i}\exp(-a/\eta)(2-y^{1/\kappa})] \} ,
\label{eq:lhal}
\end{eqnarray}
where $\xi \equiv R_\mathrm{\hal,max}/R_\mathrm{\hal}$ is set by the threshold surface
density of star formation $R_\mathrm{\hal,max} \equiv R_\mathrm{g} \ln (\Sigma_\mathrm{g,0}/\Sigma_\mathrm{th})$;
and the threshold gas surface density is adopted as
$\Sigma_\mathrm{th}\simeq 5\msun$~pc\persq (Martin \& Kennicutt 2001);
$\Sigma_\mathrm{g,0}$ is the surface density of gas at $R=0$;
$\eta=R_\mathrm{d}/R_\mathrm{\hal}$ and $\kappa=h_\mathrm{d}/h_\mathrm{\hal}$.
In order to reduce the number of free parameters ($\Sigma_{\hal,0}$, $\tau_\mathrm{\hal,0}$,
$\cos i$, $\xi$, $\eta$ and $\kappa$), we use the empirical relation between
the SFR surface density $\Sigma_{\mathrm{SFR}}$ and gas surface density
$\Sigma_\mathrm{g}$ (known as Schmidt-Kennicutt law; Kennicutt et al. 1998):
\begin{eqnarray}
\Sigma_{\mathrm{SFR}}\ (\msun \mathrm{yr\per kpc\persq}) &  \nonumber \\
= (2.5\pm0.7) & \times 10^{-4} \left(\frac{\Sigma_\mathrm{g}}{\msun
\mathrm{pc}\persq}\right)^{1.4\pm 0.15} .
\end{eqnarray}
Reminding that the gas and \hii\ distributions are both exponential, we will
get $R_\mathrm{g}=1.4R_\mathrm{\hal}$.  With above relations and the conversion
between \lhal\ and SFR (Equation~\ref{eq:sfr-lha}), we can write:
\begin{equation}
\xi = \ln\frac{\Sigma_\mathrm{\hal,0}}{4.4\times 10^{38}~\mathrm{\ergs~kpc}\persq}  .
\end{equation}

Assuming the dust-to-gas ratio to be the same as in the Galactic disc, i.e.,
$\ebv\ = 1.7 \times 10^{-22} N_\mathrm{H} (Z/\zsun)$ (Bohlin 1978)
and a Galactic extinction curve (Fitzpatrick 1999), we obtain the effective
optical depth to \hal\ through the disc perpendicularly as
\begin{equation}
\tau_\mathrm{\hal,0}=0.5\left(\frac{\Sigma_\mathrm{\hal,0}}{3.3\times10^{39}~
    {\mathrm \ergs~\mathrm{kpc}\persq}} \right)^{0.714}\frac{Z}{\zsun} .
\label{eq:tau0}
\end{equation}
Once we have $\tau$, it is straightforward to calculate the extinction at
\hal\ using Equation~\ref{eq:lhal} as following
\begin{eqnarray}
\ahal\ & = & -2.5 \log \left( \frac{L_{\mathrm{H{\alpha}},\tau}}{L_{\mathrm{H{\alpha}},\tau=0}} \right) \    \nonumber \\
       & = & \int_0^1 dy \int_0^\xi ada \exp(-a)
 \{\exp [-\frac{\tau_\mathrm{\hal,0}}{\cos i}\exp(-a/\eta)y^{1/\kappa}] +   \nonumber \\
       & & \hspace*{2em} \exp[-\frac{\tau_\mathrm{\hal,0}}{\cos i}\exp(-a/\eta)(2-y^{1/\kappa})] \} /  \nonumber \\
       & & \hspace*{3em} 2[1-\exp(-\xi)-\xi \exp(-\xi)]  ,
\label{eq:ahal}
\end{eqnarray}
Similarly, replacing $\tau_\mathrm{\hal,0}$ with $\tau_\mathrm{\hbe,0}$,
one obtains the extinction at \hbe.  Thus \hal/\hbe\ ratio can be calculated,
and then \lhal\ can be corrected with the extinction estimated from the
Balmer decrement, as the extinction correction we make to the observed \lhal.

In order to examine if such a model can reproduce the observed correlations,
we compute the average surface brightness \bhal\ and color excess \ebv\
estimated by \hal/\hbe\ from the model based on the following parameters:
$\Sigma_{\hal,0}$, $Z$, $\cos i$, $\eta$ and $\kappa$.  To facilitate
comparison with observation, we obtain \bhal\ following the same method
used for observation (Equation~\ref{eq:br}) by dividing the attenuation-corrected
\lhal\ with the surface within the half-light radius, defined as the radius
within which the \hal\ luminosity is half of the total \lhal.  The \ebv\ is
calculated from \ahal\ and \ahbe\ assuming Fitzpatrick's extinction curve.
We first fix $\eta$ to be 1.  Since increasing metallicity and inclination
both cause an increase in extinction, these two factors are coupled
in the model.  Therefore, we incorporate them into one parameter
$Zi\equiv Z/\cos i$.  We compute the models for a grid of parameters with
$38.5< \log \Sigma_{\hal,0}<42.0$, $Zi=0.5,\ 1.5,\ 2.5,\ 3.5,\ 4.5$,
$\kappa=0.5,\ 1,\ 2,\ 5$.

\begin{figure*}
\includegraphics[width=0.48\textwidth]{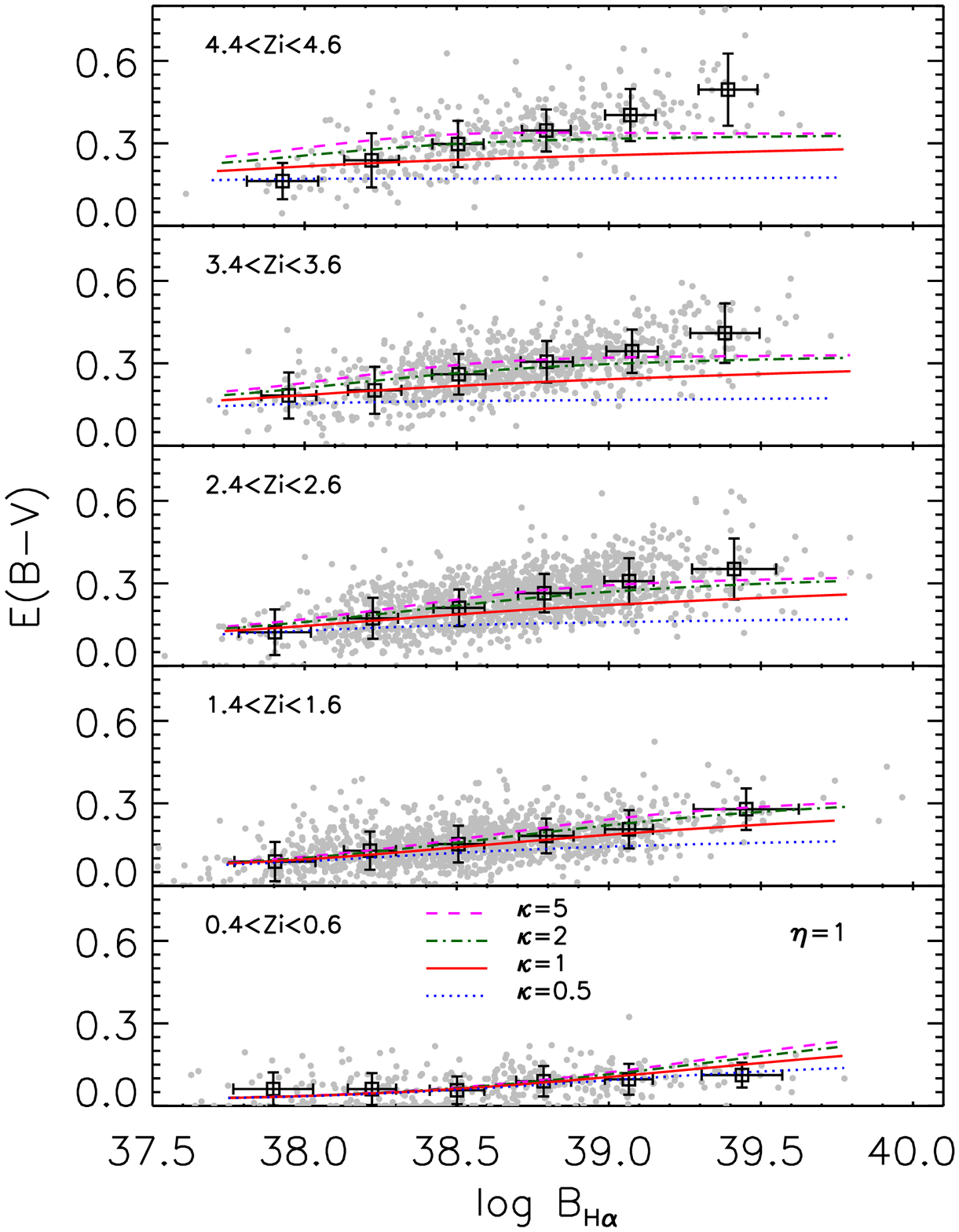}
\includegraphics[width=0.48\textwidth]{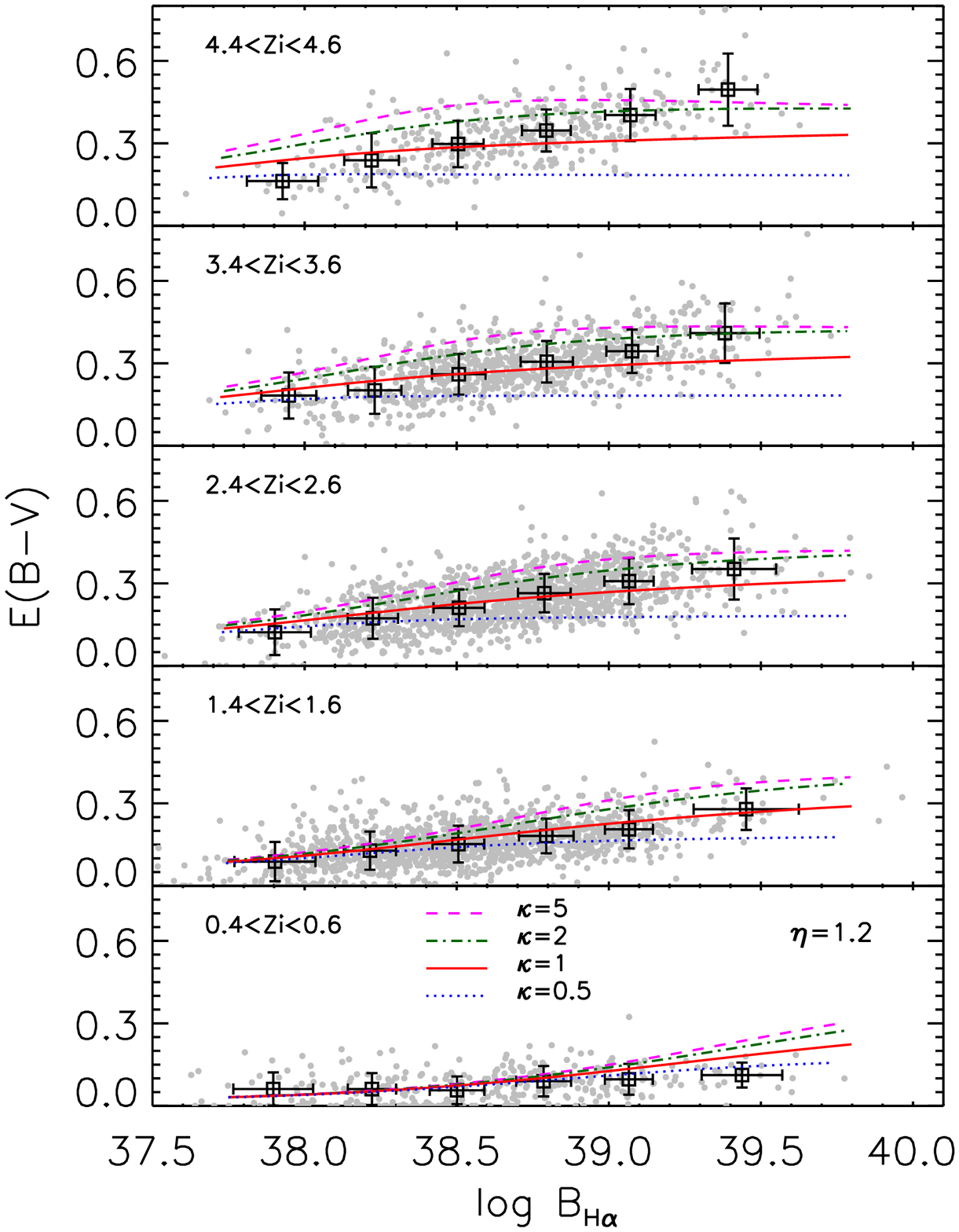}
\caption{Average surface brightness \bhal\ (\ergs~kpc\persq, corrected for
extinction) versus \ebv, with the relative characteristic lengths of dust
to \hii\ regions distribution in radial direction $\eta=1$ (left panel) and
$\eta=1.2$ (right panel).  Observation sample with $Z/\cos i$,
marked as $Zi$ for short, in bins of $0.5\pm0.1$, $1.5\pm0.1$, $2.5\pm0.1$,
$3.5\pm0.1$, and $4.5\pm 0.1$ are shown.  Dotted (blue) line represents for
the relative characteristic lengths of dust to \hii\ regions distribution in vertical direction
$\kappa=0.5$, solid (red) line for $\kappa=1$, dashed (magenta) line for
$\kappa=2$, and dash-dotted (green) line for $\kappa=5$. } \label{fig:br-ebv}
\end{figure*}

The models are over-plotted on the observation sample on the \bhal\ versus
\ebv\ plane (Figure~\ref{fig:br-ebv}).  For clarity, we only show the
observation sample in five bins of $Zi$ with a bin-width of 0.2, and on
each panel the subsample is binned in \bhal\ and shown in squares with
error bars.  We can see the observation trend on each panel, that \ebv\
increases as \bhal\ increases.  The model curves are calculated with the
corresponding value of median $Zi$ for each subsample.  On each panel,
\ebv\ gets larger when $\kappa$ increases, and each curve with constant
$\kappa$ first increases and then becomes flat as \bhal\ increases.  At
small $\kappa$, i.e. the dust layer is much thinner than the stellar disc,
\ebv\ saturates at a small value because the Balmer-line emission from the
disc of larger height tends to dominate, resulting in a small \ebv, while
at large $\kappa$ more Balmer-line emission comes from the inner region close
to the midplane of the disc, thus tend to have a large \ebv.  The saturation
is due to the fact that the observed light from outside weights more as the
optical depth increases.

Generally the $\kappa=1$ model reproduces the
observed correlations at low $Zi$.  However, as $Zi$ increases, the constant
$\kappa$ model is lower than the observation trend at the high-end
of \bhal.  At medium $Zi$ (1.5$\sim$2.5), the observation can be better
presented by the model with larger $\kappa$ (e.g. $\kappa \geq 2$).  This
could be explained by the vertical expansion of dust distribution in the
disc while higher surface brightness indicates intenser star-formation
activity.  The dust may be driven away from the midplane of the disc by
radiation pressure or the stellar winds, thus the dust layer become thicker
than that for low \bhal.  But \ebv\ saturates at about 0.3 mag when
$\kappa \geq 2$, which means that the effective optical depth saturates when
the height of dust layer gets to twice height of \hii\ regions.
Another possible explanation is that the Kennicutt-Schimidt law we
adopt in the model is an averaged star formation law that linked between
the surface density of gas and star formation as the form of
$\Sigma_{\mathrm{SFR}} \propto \Sigma_\mathrm{g}^N$.  Leroy et al. (2008) suggested
that the star formation law may vary as a function of local conditions at different
galactocentric radius, including metallicity and the dominated gas
content (H\,{\sc i} or H$_2$).  The SF law has smaller $N$ value at higher
$\Sigma_{\mathrm{SFR}}$ in the inner region of a galaxy (e.g. Bigiel et al. 2008,
their Figure 11), thus large dust reddening in the inner region dominates the observed
\ebv.  As a result, the dust reddening is higher than that the present model predicts.

In the model above, we assume $\eta=1$, i.e. the dust and \hii\ region has the same
characteristic scale in the radial distribution, but the gas/dust may
distribute at larger radii than the star-forming regions (e.g. Bigiel et al.
2008).  Thus we also consider $\eta=1.2$, and find that the observation
with large $Zi$ can be well explained by model with $\kappa=1$ at low
\bhal\ and $\kappa \geq 2$ at high \bhal.  It is possible because in a
spiral galaxy centre, where the metallicity is higher than outside, we are observing the
star-formation dominated centre, and the gas (or dust) distribution is more
extended.

\begin{figure}
\includegraphics[width=0.48\textwidth]{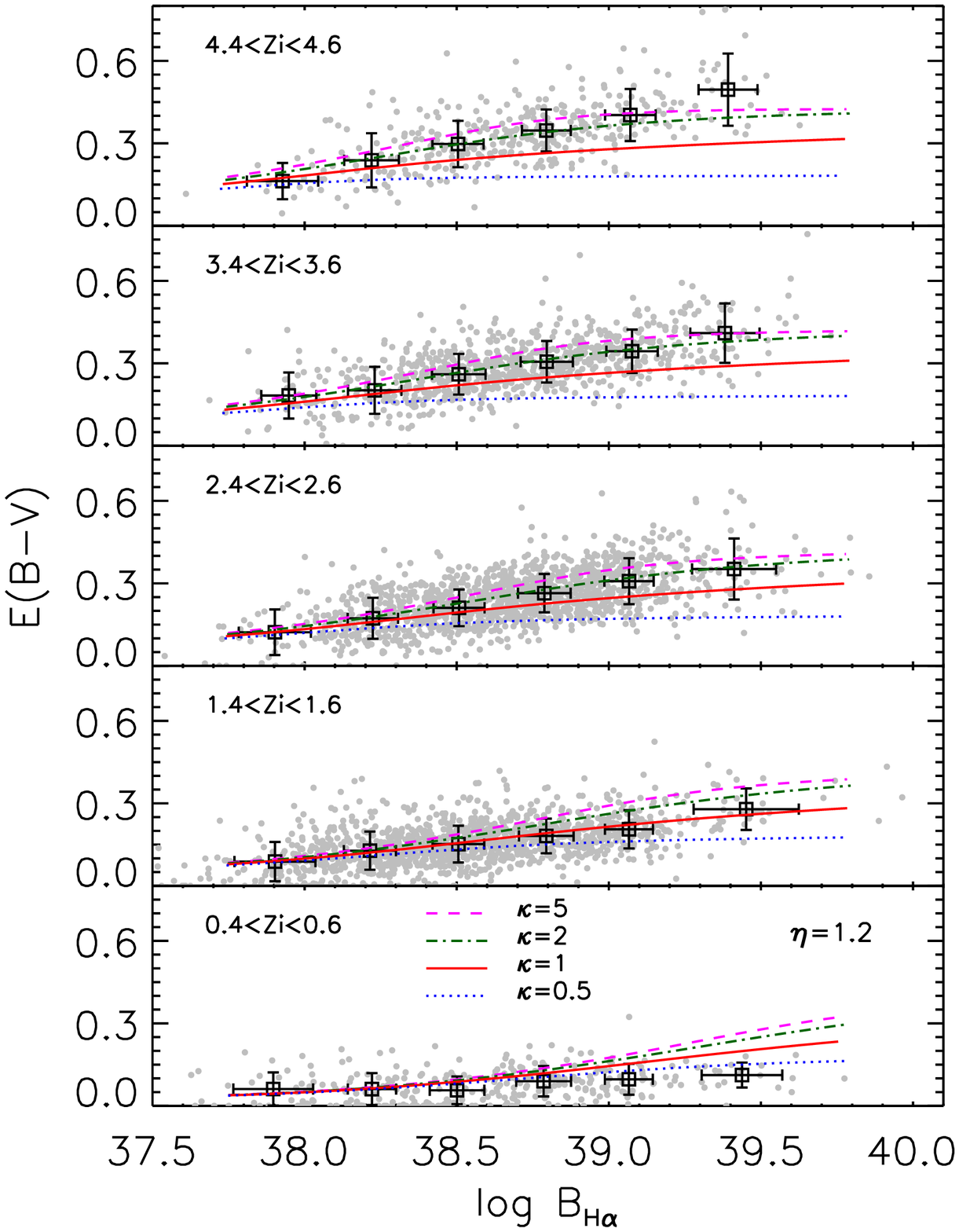}
\caption{Same figure as Figure~\ref{fig:br-ebv}, except that the model is calculated
with Equation~\ref{eq:tau0} and \ref{eq:ahal}, but adjusting
$Z/\zsun$ to $(Z/\zsun)^{0.7}$, and $\cos i$ to $(\cos i)^{0.7}$.
} \label{fig:br-ebv-z07}
\end{figure}

In this simple toy model, the optical depth $\tau_{\hal}$ is assumed to be
proportional to the metallicity $Z$.  It comes from the assumption that
$\rho_\mathrm{d}\propto\rho_\mathrm{g}\times Z$.  However, there is evidence that
this assumption may be not the case in practice.  Boissier et al. (2004)
studied six nearby late-type galaxies in FIR and UV images, and find that the
dust-to-gas ratio is proportional to $(Z/\zsun)^{0.88}$, which was flatter
than the linear correlation we adopted above.  This suggests that the
dust-to-gas ratio may not evolve linearly with $Z$. In addition, we assume
in the model that the disc height is very thin, i.e. the disc height is
quite small compared to the disc radius, thus the radial gradient in gas
density can be neglected when calculating $\tau_{\hal}$.  Under that
assumption, the effect of inclination on optical depth can be expressed
as $(\cos i)^{-1}$.  However, in practice, the disc may not have a uniform
height, for instance, it may be thicker in the centre and thinner at large radii,
or the presence of bulge in the disc centre.  Also, the distribution of
dust and star-formation regions may be not smooth, for example when
the spiral arms are present in the disc galaxy.  In those cases,
the inclination factor might be flatter than $(\cos i)^{-1}$.  Accounting
for these realistic effects, we adjust $Z/\zsun$ to $(Z/\zsun)^{\alpha}$ in
Equation~\ref{eq:tau0} and $(\cos i)^{-1}$ to $(\cos i)^{-\alpha}$
in Equation~\ref{eq:lhal} and \ref{eq:ahal}, then test several
values for the power-law index\footnote{Because the metallicity and
inclination are coupled in this simply toy model, the power-law index of
$\cos i$ is adopted to be the same as that of $Z$, so as to be incorporated
into $Zi$.}.  It turns out that with $\alpha=0.7$ the simple model reproduces
the observation trend well (Figure~\ref{fig:br-ebv-z07}).  When $Zi$ is low,
the model of constant $\kappa$ (0.5 or 1) reproduces the observation trend well.
While $Zi$ is higher, the observation can be reproduced with the model of
$\kappa \geq 2$.

To summarize, the simple toy models reproduce the observed trends between
dust reddening and \hal\ surface brightness, that reddening increases with \bhal.
The comparisons between observation and the toy models indicate that, the relative
distribution of \hii\ regions to gas/dust both in vertical and radial direction of
the galaxy disc, may vary in different environment of star-formation activity levels
or metallicities.  The dust might be distributed further beyond the \hii\ regions
when the star formation activity is intenser, or the dust is distributed in a thicker
disc ($\kappa \geq 1$) than the \hii\ regions when the metallicity gets higher.
In addition, our results imply that dust-to-gas density ratio may be not proportional
to the metallicity $Z$, but to a power-law of $Z$, $Z^{\alpha}$ with $\alpha\sim0.7$.

\section{CONCLUSIONS AND IMPLICATIONS}
\label{sec:conclusion}

We present the empirical formulae of intrinsic reddening as a function
of intrinsic \hal\ luminosity or surface brightness, metallicity and disc
inclination (Equation~\ref{eq:powfit}, \ref{eq:powfit_br} and
Table~\ref{tab:lha} and \ref{tab:br}) for a large sample of $\sim$22 000
well-defined star-forming disc galaxies selected from the SDSS.  With the
empirical formulae, the reddening parameterized by \ebv\ could be predicted
within 1$\sigma$ uncertainty of 0.07 mag.  The \debv, defined as the
observed reddening estimated from the Balmer decrement minus predicted
value, does not correlate with the three parameters used in the formulae.
We also find that \debv\ is independent of the stellar mass, 4000 \AA\ break
strength or electron density within the galaxy.

The observed trend between the dust reddening and \hal\ surface
brightness could be reproduced by a plane-parallel slab toy model,
in which the dust is scaled with gas, and all the distributions of
the dust, gas and star-formation regions are smooth and follow exponential
laws in the disc.  We find from the comparisons of the models to the
observation, the relative vertical scale of dust distribution to \hii\ regions
distribution may vary with \hal\ surface brightness or metallicity.
The higher intensity of star formation could drive the dust to distribute
in a thicker layer, or the dust disc is thicker than the disc of the \hii\
regions when the metallicity is higher.  However, the real galaxies may be
not fully described by this simple toy model.  For example, when the spiral arms
are present, or the SF regions are clumpy, the assumptions of smooth distributions
will be improper.  In addition, our result implies a different scaling law of
dust-to-gas ratio as a function of metallicity $Z$ from the linearly
relation (Equation~\ref{eq:rhod1}).  Our empirical relations, which suggest
dust reddening is partially dependent on metallicity ($\propto Z^{0.6\sim0.7}$),
can be an observational constraints on the output of radiative-transfer model
calculations, and hence provides constraints on the assumed dust-to-gas ratio in the model.

Since metallicity-dependence has been introduced, we believe that our
empirical formulae should be applicable to the early evolutionary stage
of a galaxy, when the gas metallicity might be low.  In the starburst
galaxies at high redshift, dust surrounding star-forming region has been
proven to be prone to sub-milimeter and far-infrared observations.  But most
of those galaxies are relatively dim in optical, implying severe
extinction and reddening to these galaxies.  The observed galaxies
at high redshifts are affected by reddening, in the way that the galaxies
with heavy attenuation and reddening may escape from the detection.
The importance of the reddening correction depends on the wavelength
band that is used.  For example, UV continuum luminosity is usually
used to infer SFR in high-redshift galaxies.  It is very sensitive
to the attenuation/reddening.  Calzetti et al. (2000) find that the
extinction to emission lines and to stellar continuum in optical band
are different, and the continuum reddening is only about 40 percent of that
for emission lines.  However, the extinction to UV continuum is
likely similar to the \hii\ region because UV continuum is also
emitted by young hot stars, which are likely surrounded by \hii\
regions.  As a result, our results may be used to estimate the reddening
for these high-redshift galaxies with caution, that the metallicity should
be well-calibrated and be within the range covered by our sample.  Following
Equation~\ref{eq:powfit} and the discussions in Section~\ref{sec:relation},
SFR in luminous star formation or high metallicity galaxies might have been
under-estimated (e.g. Heckman et al. 1998; Hopkins et al. 2001; Panuzzo et al. 2007),
or those galaxies are even completely lost in optical surveys.

The quantitative relationship of reddening with the metallicity, luminosity
or surface brightness, and the inclination for disc galaxies can also be
incorporated into the current semi-analytic models of galaxy formation and
evolution (e.g. Croton et al. 2006; De Lucia et al. 2004; Kang et al. 2005).
In those models, a simple chemical enrichment scheme has been included, and
the metal abundance has been actually predicted.  With the known metallicity
and the star formation rate (intrinsic luminosity) provided by the models,
the reddening can be estimated accurately with our formulae, and then used to
shape the spectral energy distribution and predict the emerging luminosity
more realistically.

\section*{Acknowledgments}

We would like to thank the anonymous referee for useful suggestions that
improved the paper.  We thank Gleniese Mckenzie, Peng Jiang for helpful
comments on the manuscripts.  This work is supported by
NSFC (10973013, 11073017, 11033007), 973 program (2009CB824800) and
the Fundamental Research Funds for the Central Universities.
Funding for the SDSS and SDSS-II has been provided by the Alfred P. Sloan
Foundation, the Participating Institutions, the National Science Foundation,
the U.S. Department of Energy, the National Aeronautics and Space Administration,
the Japanese Monbukagakusho, the Max Planck Society, and the Higher Education
Funding Council for England. The SDSS Web Site is http://www.sdss.org/.  The SDSS
is managed by the Astrophysical Research Consortium for the Participating
Institutions. The Participating Institutions are the American Museum of
Natural History, Astrophysical Institute Potsdam, University of Basel,
University of Cambridge, Case Western Reserve University, University of
Chicago, Drexel University, Fermilab, the Institute for Advanced Study,
the Japan Participation Group, Johns Hopkins University, the Joint Institute
for Nuclear Astrophysics, the Kavli Institute for Particle Astrophysics and
Cosmology, the Korean Scientist Group, the Chinese Academy of Sciences (LAMOST),
Los Alamos National Laboratory, the Max-Planck-Institute for Astronomy (MPIA),
the Max-Planck-Institute for Astrophysics (MPA), New Mexico State University,
Ohio State University, University of Pittsburgh, University of Portsmouth,
Princeton University, the United States Naval Observatory, and the
University of Washington.

\clearpage

\label{lastpage}

\end{document}